\documentclass{JHEP3}

\newcommand{\half}{{{\textstyle\frac{1}{2}}}}
\newcommand{\quarter}{{{\textstyle\frac{1}{4}}}}

\newcommand{\be}{\begin{equation} }
\newcommand{\ee}{\end{equation} }
\newcommand{\ba}{\begin{array}}
\newcommand{\ea}{\end{array}}

\newcommand{\bea}{\begin{eqnarray}}
\newcommand{\eea}{\end{eqnarray}}

\newcommand{\nn}{\nonumber}

\def\a{{\alpha^{\prime}}}

\def\tr{{\rm tr}}
\def\Tr{{\rm Tr}}

\title{Thermodynamic behavior of IIA string theory on a pp-wave}

\author{Seungjoon Hyun${}^{\ast}$, Jong-Dae Park$^{\dagger\ddag}$ and Sang-Heon
Yi$^{\ast}$\\
${}^{\ast}$Institute of Physics and Applied Physics, Yonsei University, Seoul 120-749, Korea\\
${}^{\dagger}$Yonsei Visiting Research Center, Yonsei University,
Seoul 120-749, Korea\\
${}^{\ddag}$Center for Theoretical Physics, Seoul National
University, Seoul  151-742, Korea\\

E-mail~:~\email{hyun@phya.yonsei.ac.kr},
\email{jdpark@phya.snu.ac.kr},
 \email{shyi@phya.yonsei.ac.kr}}

\abstract{We obtain the thermal one loop free energy and the
Hagedorn temperature of IIA superstring theory on the pp-wave
geometry which comes from the circle compactification of the
maximally supersymmetric eleven dimensional one. We use both
operator and path integral methods and find the complete agreement
between them in the free energy expression. In particular,  the
free energy in the $\mu \rightarrow \infty$ limit is shown to be
identical with that of IIB string theory on maximally
supersymmetric pp-wave, which indicates the universal thermal
behavior of strings in the large class of pp-wave backgrounds. We
show that the zero point energy and the modular properties of the
free energy  are naturally incorporated into the path integral
formalism.  }


\preprint{hep-th/0304239 \\
          SNU-TP 03-009}

\begin{document}

\section{Introduction and conclusion}

String/M theory is often considered as the theory of everything.
In particular, it is expected to describe the early
Universe~\cite{Polchinski:rq}. In this sense, it is very important
to understand the finite temperature physics of the string/M
theory. As is well known, there are some difficulties in studying
the thermodynamics of a theory which contains gravity, for example
string theory. Gravity suffers from the Jeans' instability which
is deeply related to the formation of black holes. Also, the
energy of a system with gravity may be difficult to define. But,
as was argued in~\cite{atickwitten}, we may consider the weakly
coupled low energy limit of such theory and get some insight on
the thermodynamic behavior of the theory. The most notable feature
in the finite temperature string theory is the existence of the
Hagedorn temperature~\cite{Hagedorn:st,huang} around the string
scale due to the exponential growth of the level density of string
states. This Hagedorn temperature is believed to be the signal of
a phase transition~\cite{atickwitten}, though there is no clear
understanding on the physics above the Hagedorn temperature.

There were some interesting
works~\cite{PandoZayas:2002hh,Greene:2002cd,Sugawara:2002rs,
Brower:2002zx,Sugawara:2003qc,semenoff} on the finite temperature
 IIB superstring theory on the maximally
supersymmetric ten-dimensional pp-wave spacetime~\cite{bla242}
which can be obtained as the Penrose limit of the $AdS_5 \times
S^5$ spacetime and gives another solvable model for Green-Schwarz
superstrings~\cite{Metsaev:2001bj}. As one incarnation of AdS/CFT
correspondence~\cite{Berenstein:2002jq}, the finite temperature
behavior of IIB string theory on this pp-wave has been studied in
connection with that of the dual ${\cal N} =4$ super Yang-Mills
theory in the corresponding limit. In those works, one loop free
energy is calculated and  the dependency of the Hagedorn
temperature on the RR-flux, $\mu$, is found. If we assume that the
Hagedorn temperature of strings on the pp-wave geometry depends on
$\mu$ smoothly so that it reduces to the flat space result in the
$\mu\rightarrow 0$ limit, then, from the dimensional argument, it
can be expanded in terms of dimensionless quantity
${\sqrt{\a}}\mu$ as
\be T_H = \frac{1}{2\pi\sqrt{2\a}}\Big(1 + \sum_{n=1}^\infty
a_n(\sqrt{\a}\mu)^n \Big) \,, \label{hagedorn}\ee
where $a_n$ is the numerical constant depending on the details of
pp-wave geometry. It was argued in \cite{Greene:2002cd} that this
Hagedorn temperature in the case of IIB pp-wave geometry represents
the limiting temperature of string theory where
the free energy diverges. Soon after it was shown in~\cite{Brower:2002zx}
that the free energy does not diverge, but is singular, and hence the
Hagedorn temperature represents the phase transition.

In this paper, we aim to study the thermodynamics of type
IIA Green-Schwarz (GS) superstring on the ten dimensional pp-wave
geometry~\cite{hyunandshin1},
\bea ds^2 &=& -2 dx^{+} dx^{-} - A(x^I) dx^{+}dx^{+} +
\sum_{I=1}^{8 }dx^{I}dx^{I}\,, \nn \\ && \nn \\
     && F_{+123} = \mu\,, \qquad F_{+4}{}={}
     -{\textstyle\frac{\mu}{3}}\,,
\label{pp-wave}\eea
where
$$ A = ({\textstyle\frac{\mu}{3}})^2(x_1^2+x_2^2+x_3^2+x_4^2)+
  ({\textstyle\frac{\mu}{6}})^2(x_5^2+x_6^2+x_7^2+x_8^2)\,.
$$
This geometry comes from the dimensional reduction of the
maximally supersymmetric eleven dimensional pp-wave~\cite{fig308},
and has one characteristic mass scale, $\mu$. In studying the
thermal properties of IIA string theory on this pp-wave geometry,
we use the canonical ensemble. The canonical ensemble may not be
justified in the cases where the formation of the long string can not be
ignored, as its fluctuations make the thermalization of the string
gas difficult. In this case, the microcanonical ensemble should be
used instead~\cite{Frautschi:1971ij,Carlitz:1972uf,Bowick:az,
Bowick:1989us,Deo:1988jj}. It would be interesting to study the
thermodynamics of IIA string theory on pp-wave using the
microcanonical ensemble and to see the implications.

There are two complementary approaches to calculate the one loop
free energy of the finite temperature string theory. One is the
operator method which can be adapted easily once the light-cone
gauge is chosen. The other is the path integral approach which is
conceptually more transparent and is ready to be generalized to higher
loops. In this paper we compute the one loop free energy using
both methods and find the perfect agreement. Though, on general
grounds, this should be expected, the explicit confirmation is rather
nontrivial. One issue in the path integral approach to compute
thermal one loop partition function is to determine the correct
path integral measure, which is settled down in this paper.

We also obtain the Hagedorn temperature which takes the assumed
form (\ref{hagedorn}). It is found to be the monotonic function of
${\sqrt \a}\mu$ and goes to infinity as $\mu\rightarrow \infty$.
In the same limit, the free energy is shown to become
\be F = -\frac{L\pi}{6\beta^2}\,, \ee
which is exactly the same as that of IIB string theory on the
pp-wave~\cite{semenoff}. In general, the free energy of $d$
dimensional field theories at high temperature behaves
as~\cite{atickwitten}
\be F_d \sim V \beta^{-d}\,. \ee
This suggests that in the large $\mu$ limit, the physics on the
pp-wave geometry essentially reduces to two-dimensional one. It
would be very interesting to pursue this line.

 As will be shown later, the one loop free
energy of IIA superstrings on the given pp-wave geometry has very
similar structure to the IIB counterpart, apart from the small
modification due to the differences in the supersymmetry and their
supermultiplets. This strongly suggests that the Hagedorn
temperature signals the phase transition, like
 the IIB case~\cite{Brower:2002zx}.

Our main motivation for the present work is to understand the real
degrees of freedom of IIA strings on pp-wave and eventually
to reveal those of M theory on pp-wave.  The eleven-dimensional
maximally supersymmetric pp-wave can be obtained from the Penrose
limit of $AdS_4\times S^7$ or $AdS_7\times S^4$. Since the M
theory on those $AdS$ spacetimes are dual to three dimensional or
six dimensional conformal field theories, respectively, it is
natural to expect that, in the spirit of the AdS/CFT
correspondence, the M theory on the pp-wave is dual to (the common
sector of ) those theories in the corresponding limit. Our study
on the finite temperature physics of IIA strings on pp-wave may
give some clues to understand those dual theories.

This paper is organized as follows. In section two, we briefly
review IIA string theory on the pp-wave geometry. In section
three, we obtain the free energy of the weakly interacting string
gas on the pp-wave using the operator method. In section four, we
consider the asymptotics of the free energy and find the Hagedorn
temperature which depends on the RR-flux through the difference of
boson and fermion zero point energies. In section five, we use the
path integral approach to calculate the one loop free energy.
Using this approach, we clarify several issues, including the zero
point energy and the modular properties of the one loop free
energy. In section six, we show the complete agreement between the two
results, one from the operator method and the other from the path
integral method.
We also give
various comments including the correct thermal free energy of the
GS strings on flat space. In the appendices, we give simple
derivation of the one loop free energy using Coleman-Weinberg
formula and also give some useful formulae used in the main text.

\section{Review: IIA superstring on the pp-wave geometry}

In this section we summarize some results for the type IIA
superstring theory on the ten dimensional pp-wave spacetime,
including the mode expansions, the normal ordered worldsheet
Hamiltonian and momentum (see, for more
details,~\cite{hyunandshin1,hyunandshin2}). The IIA pp-wave
geometry (\ref{pp-wave}) admits 24 Killing spinors, as the torus
compactification breaks 8 supersymmetries out of 32 in the
eleven-dimensional pp-wave geometry. In the light-cone gauge, the
worldsheet description of IIA Green-Schwarz superstrings on this
pp-wave background is given by the two-dimensional, free massive
${\cal N}=(4,4)$ supersymmetric theory. The bosonic fields, the
target space coordinates, split into two multiplets with masses
proportional to $\frac{\mu}{3}$ and $\frac{\mu}{6}$ in the
light-cone gauge. The fermionic superpartners, after kappa gauge
fixing, also split into two multiplets with the same masses as
bosonic partners, classified by the chirality of $SO(8)$ rotations
in $x^I$ and $SO(4)$ rotations in $x^{i'}$\footnote{In what
follows, the indices $I=1,2,...,8$ denote the 8 transverse
coordinates, among which the indices $i=1,...,4$ denote the
direction where RR field strengths are spanned and the indices
$i^{\prime}=5,...,8$ denote the other directions. We also take the
real and symmetric representation for $SO(8)$ gamma matrices. }.
This is due to the fact that eight Killing spinors, which
correspond to the dynamical supersymmetry on the string
worldsheet, do not explicitly depend on the coordinate $x^+$. Let
us denote the chirality of $SO(8)$ and $\gamma^{1234}$ as
superscripts $1,2$ and subscripts $\pm$ for fermions,
$\psi_{\pm}^{1,\,2}$,  respectively. Then one can see that the
theory contains two supermultiplets $( X^i, \psi_-^1, \psi_+^2 )$
and $( X^{i'}, \psi_+^1,
  \psi_-^2 )$ whose masses are given by $\frac{m}{3}$ and
$\frac{m}{6}$, respectively, with $ m \equiv \mu\a p^{+}$~.

The light-cone gauge fixed action for this IIA Green-Schwarz
superstring becomes quadratic in its fields as
\bea S_{LC} &=& -\frac{1}{4\pi\a}\int d\sigma^{0}d\sigma^{1}
\Big[\partial_{\alpha}X^I\partial^{\alpha}X^I +
({\textstyle\frac{m}{3}})^2(X^i)^2+
  ({\textstyle\frac{m}{6}})^2(X^{i'})^2
   -i\psi^1_{-}\partial_{+}\psi^1_{-}
   -i\psi^2_{+}\partial_{-}\psi^2_{+}\nn \\ && \nn \\
&&{}~~~~~~~~~~~~~~~~~~~{}
  -i\psi^1_{+}\partial_{+}\psi^1_{+}
  -i\psi^2_{-}\partial_{-}\psi^2_{-}
   + i\frac{2m}{3}\psi^2_{+}\gamma^4\psi^1_{-}
   -i\frac{m}{3}\psi^2_{-}\gamma^4\psi^1_{+}\Big]\,,
\eea
where $\alpha=0,1$ denotes worldsheet index and
$\partial_{\pm}=\partial_{0}\pm\partial_{1}$.

Each field in the multiplet $( X^i, \psi_-^1, \psi_+^2 )$ can be
mode expanded, in terms of frequencies
\be
    n\ge 0 \quad ; \quad
 \omega_n =  \sqrt{ ( {\textstyle\frac{m}{3}})^2 +
 n^2}\,,\qquad
 n<0  \quad ; \quad
  \omega_n =  - \sqrt{ ( {\textstyle\frac{m}{3}})^2 + n^2
    }\,,  \ee
as follows:
\bea X^{i}
&=&i{\textstyle\sqrt{\frac{\a}{2}}\sqrt{\frac{1}{\omega_0}}}
 {}~{} \Big(a^{i}e^{-i\omega_0\sigma^0} -a^{i\,
 \dagger}e^{i\omega_0\sigma^0}\Big) +
i{\textstyle\sqrt{\frac{\a}{2}}}\sum_{n\neq 0}
\Big({\textstyle\frac{1}{\omega_n}}\alpha^{i}_{n}
e^{-in\sigma^{1}} +{\textstyle\frac{1}{\omega_n}}
{\tilde\alpha}^{i}_{n} e^{in\sigma^{1}} \Big)
e^{-i\omega_{n}\sigma^{0}}\,,
\nn \\ && \nn \\
\psi^1_{-}  &=& -i{\textstyle\sqrt{\frac{\a}{2}}} \gamma^4
\Big(\chi\, e^{-i\omega_0\sigma^0}- \chi^\dagger\,
e^{i\omega_0\sigma^0}\Big) + \sum_{n\neq 0} c_n\Big(
\tilde{\psi}_n{}~{} e^{in\sigma^1} -i
{\textstyle\frac{\omega_n-n}{\omega_0}}\gamma^4\psi_n{}~{}
e^{-in\sigma^1}\Big) e^{-i\omega_n\sigma^0 }\,,
\nn \\ && \nn \\
\psi^2_{+}  &=& {\textstyle\sqrt{\frac{\a}{2}}}  \Big(\chi\,
e^{-i\omega_0\sigma^0} + \chi^\dagger\, e^{i\omega_0\sigma^0}\Big)
+ \sum_{n\neq 0} c_n\Big( \psi_n{}~{} e^{-in\sigma^1} +i
{\textstyle\frac{\omega_n-n}{\omega_0}}\gamma^4\tilde{\psi}_n{}~{}
e^{in\sigma^1}\Big) e^{-i\omega_n\sigma^0 }\,, \nn\eea
where $c_n =
\sqrt{\frac{\a\omega_0^2}{\omega_0^2+(\omega_n-n)^2}}$ is the
normalization constant taken canonically. After the canonical
quantization, the modes satisfy  the following commutation
relations,
\[  [ a^i, a^{j \dagger} ] = \delta^{ij}\,, \qquad
\big[\alpha^{i}_{n}, \alpha^{j}_{m} \big] = \omega_n \delta^{ij}
\delta_{n+m,0}\,, \]

\[ \{ \chi, \chi^\dagger \} = 1 \,, \qquad  \{ \psi_n , \psi_m \} =
\delta_{n+m,0} \,,  \qquad \{ \tilde{\psi}_n, \tilde{\psi}_m \} =
\delta_{n+m,0}\,.\nn
 \]
 Each field in the
multiplet $( X^{i'}, \psi_+^1, \psi_-^2 ) $ can also be mode
expanded  and canonically quantized similarly with  frequencies
\be
 n\ge 0 \quad ; \quad \omega'_n =\sqrt{ ({\textstyle\frac{m}{6}})^2 + n^2 }\,,
  \qquad n<0  \quad ; \quad
  \omega'_n = -
      \sqrt{ ({\textstyle\frac{m}{6}})^2 + n^2 }\,.\nn
\ee
Introducing the number operators $N^{B}_{i\, n}, N^{F}_{n}$ for
the multiplet $( X^{i}, \psi_-^1, \psi_+^2 ) $, defined by
\bea  \left. \ba{cccc} & n >0 &  \qquad n=0 & \qquad n <0
\\ & & &  \\
N^{B}_{i\, n} := & \frac{1}{\omega_n}{}~{} \alpha^{i}_{-n}
\alpha^{i}_n\,, &  \qquad a^{i \dagger} a^i\,, & \qquad
 \frac{1}{~\, \omega_{-n}}{}~{} {\tilde\alpha}^i_{n} {\tilde\alpha}^i_{-n}\,,
 \\ & & &  \\
N^{F}_{n}:= &  \psi_{-n} \psi_n\,, & \qquad \chi^\dagger \chi\,, &
\qquad {\tilde\psi}_{n}{\tilde\psi}_{-n}\,, \ea\right. \eea
and similarly $N^{B}_{i'\, n}, N^{\prime F}_{n}$ for $( X^{i'},
\psi_+^1, \psi_-^2 ) $, the worldsheet momentum and Hamiltonian
can be written as
\be {\cal P}= N-\tilde{N} = \sum_{n=-\infty}^{\infty}n
\Big(\sum_{i=1}^4 N^B_{i\, n} +\sum_{i^{\prime}=5}^8
N^{B}_{i^{\prime}\, n}+ N^{F}_{n} + N^{\prime\, F}_{n} \Big)\,,\ee
\be {\cal H} = \a p^{+} p^{-} =
\sum_{n=-\infty}^{\infty}\bigg[|\omega_n|\Big(
\sum_{i=1}^4N^{B}_{i\, n} + N^{F}_{n}\Big)
+|\omega_n^{\prime}|\Big(\sum_{i^{\prime}=5}^8 N^B_{i^{\prime}\,
n} + N^{\prime\,F}_{n}\Big) \bigg]\,. \label{lcham}\ee
%

%%%%%%%%%%%%%%%%%%%%%%%%%%%%%%%%%%%%%%%%%%%%%%%%%%%%%%%%%%%%%%%%%

\section{Operator method}
In this section we use the operator method to study the
thermodynamic properties of IIA strings on the pp-wave. We
consider the canonical ensemble composed of an ideal gas of weakly
interacting strings on the given background. We mainly focus on
the thermodynamic partition function or the free energy of string
gas, from which all the other thermodynamic quantities can be
obtained.

The free energy of the string gas can be expressed in terms of the
trace, $\Tr^{\prime}$, over the one-string physical states as
 \bea F&=&\frac{1}{\beta} \Tr^{\prime}\, \bigg[ (-1)^{\bf F} \ln
\Big(1- (-1)^{\bf F} e^{-\beta p^0}\Big)\bigg]~, \nn \\ && \nn \\
&=& - \sum_{l=1}^{\infty} \frac{1+(-1)^l}{2l\beta}\,
\Tr^{\prime}\,\Big[(-1)^{\bf F}\,e^{-l\beta p^0}\Big]
-\sum_{l=1}^\infty
\frac{1-(-1)^l}{2l\beta}\,\Tr^{\prime}\,e^{-l\beta p^0}
 \,, \eea
where ${\bf F}$ is the space-time fermion number
operator\footnote{One may use $ap^{+}+bp^{-}$ instead of $\beta
p^0$ for the Boltzman factor
\cite{Greene:2002cd,Brower:2002zx,semenoff}, which is connected
with the grand canonical ensemble. The results for this
substitution can be easily read from our results as the geometry
is invariant under the boosting along the longitudinal direction
with the rescaling of $\mu$.}. The level matching condition $N =
\tilde{N}$ can be implemented by introducing the Lagrange
multiplier, $\tau_1$, as
\bea F &=& -\sum_{l=1}^{\infty} \frac{1+(-1)^l}{2l\beta}
 \Tr {}~{} \bigg[ \int_{-1/2}^{1/2}d\tau_1{}~{}  e^{2\pi i
\tau_1(N-\tilde{N})}{}~{}(-1)^{\bf F} e^{-l\beta p^{0}} \bigg]
 \nn \\
&&{}~{} -\sum_{l=1}^\infty \frac{1-(-1)^l}{2l\beta}\, \Tr
{}~{}\bigg[ \int_{-1/2}^{1/2}d\tau_1{}~{} e^{2\pi i
\tau_1(N-\tilde{N})}{}~{} e^{-l\beta p^{0}} \bigg]\,,
\label{free1} \eea
where $\Tr$ denotes the trace over the one-string Fock space
states without the level matching condition.

In the light-cone formalism, which is particularly suitable for
the given pp-wave geometry, the space-time energy
$p^{0}=\frac{1}{\sqrt{2}}(p^{+}+p^{-})$ is split into the
light-cone Hamiltonian, $p^{-}=\frac{1}{\a p^+}{\cal H}$, and
kinematical momentum, $p^{+}$. In this formalism, the one-string
states are represented by continuous $p^+$ and the transverse
oscillators. Thus the above $\Tr$ can be written as
\be \Tr = \frac{L}{\sqrt{2}\pi}\int_0^\infty dp^{+}{}~~{}
\tr_{trans}\,, \ee
where $L$ is the (infinite) length of the longitudinal
($x^{9}=\frac{1}{\sqrt{2}}(x^{+}-x^{-})$) direction and
$\tr_{trans}$ denotes the trace for the transverse oscillators
without level matching constraint.

The first term in the free energy expression in (\ref{free1}) is
simply the integral over $p^+$ of the Witten index:
\[
\tr_{trans} \Big[ \int_{-1/2}^{1/2}d\tau_1{}~{} e^{2\pi i
\tau_1(N-\tilde{N})}{}~{}(-1)^{\bf F} e^{-\frac{\beta}{\sqrt{2}}
p^{-}}\Big]~.
\]
This Witten index, in our case, turns out to be unity, while it
vanishes in the case of string theory on flat spacetime, due to
fermion zero modes.

Let us define
\be \tau_2 := \frac{1}{\sqrt{2}} \frac{l\beta}{2\pi\alpha^\prime
p^{+}}\,, \ee
in terms of which the mass parameter, $m$, can be rewritten as
\be
 m =  \frac{l\beta\mu}{2\sqrt{2}\pi\tau_2}\,. \label{mmod}\ee
This change of variables leads to
\be F = - \frac{L\pi}{24\beta^2} - \sum_{l=1,\,{\rm odd}}^\infty
\frac{L}{4\pi^2\a}\int^\infty_0 \frac{d\tau_2}{\tau^{~2}_2}{}~{}
e^{-\frac{l^2\beta^2}{4\pi\a\tau_2}} \int_{-1/2}^{1/2} d\tau_1
{}~{} \tr_{trans}\, \Big[ e^{2\pi i\tau_1{\cal P}}{}~{}
e^{-2\pi\tau_2\, {\cal H}} \Big] \,. \ee
Note that ${\cal H}=\a p^{+}p^{-}$ given in Eq.~(\ref{lcham}) has
$l$-dependence through $m$.

The integrand, $\tr_{trans}\,[ e^{2\pi i\tau_1{\cal P}}{}~{}
e^{-2\pi\tau_2\, {\cal H}}]$, can be easily computed using the
following formulae
\bea \!\!\!\!\! \prod_{i=1}^{4}\sum_{N^{B}_{i\, n}=0}^{\infty}
e^{(-2\pi\tau_2|\omega_n|+ 2\pi i \tau_1 n )N^{B}_{i\, n}} &=&
\bigg(\frac{1}{1-e^{-2\pi\tau_2|\omega_n| + 2\pi i\tau_1
n}}\bigg)^4\,, \qquad {\rm for ~~ bosons} \,,  \\ \!\!\!\!\! && \nn \\
\!\!\!\!\! \sum_{N^{F}_{n}=0}^{4}{4\choose N^F_n}
e^{(-2\pi\tau_2|\omega_n| + 2\pi i \tau_1 n) N^{F}_{n}} &=& \Big(1
+ e^{-2\pi\tau_2|\omega_n| + 2\pi i\tau_1 n}\Big)^4 \,, \qquad
{\rm for ~~ fermions} \,, \eea
with  similar ones for $N^{B}_{i'\, n}, N^{\prime F}_{n}$. As a
result, the free energy becomes
\bea  F &= &- \frac{\pi L}{24\beta^2} {}~{} - \sum_{l=1,{\rm
odd}}^\infty \frac{L}{4\pi^2\a}\int^\infty_0
\frac{d\tau_2}{\tau^{~2}_2} \int^{1/2}_{-1/2} d\tau_1 {}~{}
e^{-\frac{l^2\beta^2}{4\pi\a\tau_2}} \nn \\ &&\nn \\ &&
{}~~~~~{}~~~~~{} \prod_{n=-\infty}^\infty \left(\frac{1+ e^{-2\pi
\tau_2|\omega_n| + 2\pi i \tau_1 n } } {1-e^{-2\pi
\tau_2|\omega_n| + 2\pi i \tau_1 n }}\right)^4 \left(\frac{1+
e^{-2\pi \tau_2|\omega^\prime_n| + 2\pi i \tau_1 n } } {1-e^{-2\pi
\tau_2|\omega^\prime_n| + 2\pi i \tau_1 n }}\right)^4\,.  \eea
Since we have used the normal ordered Hamiltonian, the zero point
energy contribution to the Hamiltonian cancels between bosons and
fermions and does not appear in the above free energy expression.

\section{Free energy and Hagedorn temperature}

In this section, we obtain the Hagedorn temperature from the free
energy given in the previous section. We will see that the
Hagedorn temperature behaves smoothly with respect to RR-flux,
$\mu$, and in fact is a monotonic function of $\mu$.

The Hagedorn temperature can be extracted from the asymptotic
behavior of the free energy. In order to see its asymptotic
behavior, it is convenient to introduce the following quantity:
\be D_{b_1,\, b_2}(\tau_1,\tau_2; m)  :=
e^{2\pi\tau_2\Delta_{b_1}(m)}\prod_{n=-\infty}^\infty \Big(1 -
e^{-2\pi\tau_2\sqrt{(n+b_1)^2+m^2} + 2\pi i \tau_1(n+b_1) - 2\pi i
b_2}\Big) \,, \label{Dbb}\ee
where $\Delta_b(m)$ is defined by
\be \Delta_{b}(m) :=-\frac{m}{\pi}\sum_{p=1}^\infty
\frac{\cos(2\pi b p)}{p}K_1(2\pi m p) =
-\frac{1}{2\pi^2}\sum_{p=1}^\infty \cos(2\pi b p){} \int_0^\infty
d s {}~{}  e^{-p^2 s -\frac{\pi^2 m}{s}}\,, \label{delta} \ee
with the modified Bessel function $K_1(x)$. The function
$\Delta_b(m)$ can be identified as the physical zero point (or
Casimir) energy for two dimensional massive boson or fermion field
with the twisted boundary condition as will be shown in the path
integral approach given in the next section.
 Note that
\be  \Delta_{b} (m) {}~{} {\buildrel m\rightarrow
0\over\longrightarrow} {}~~~{}\frac{1}{24}-\frac{1}{8}(2b-1)^2\,,
\qquad  \Delta_{b} (m) {}~{}{\buildrel m\rightarrow
\infty\over\longrightarrow}{}~~~{} 0\,.\ee

   We can
exhibit the free energy using $D_{a,\, b}(\tau_1,\tau_2; m)$ as
\bea
 F &=& - \frac{\pi L}{24\beta^2}
 - \sum_{l=1\atop~~{\rm odd}}^\infty \frac{L}{4\pi^2\a}
\int^\infty_0 \frac{d\tau_2}{\tau^{~2}_2}
\int^{\half}_{-\half}d\tau_1
{}~~{}e^{-\frac{l^2\beta^2}{4\pi\a\tau_2}} \nn\\ && \nn \\
&& {}~~~~~~~~~~~~~~~~~~~~~~~~~~~{} \times ~
\bigg[\frac{D_{0,\,1/2}(\tau_1,\tau_2;{\textstyle\frac{m}{3}})}
{D_{0,\,0}(\tau_1,\tau_2;{\textstyle\frac{m}{3}})}\bigg]^4
\bigg[\frac{D_{0,\,1/2}(\tau_1,\tau_2;{\textstyle\frac{m}{6}})}
{D_{0,\,0}(\tau_1,\tau_2;{\textstyle\frac{m}{6}})}\bigg]^4 \,.
\label{Free}\eea
Since $D_{0,\, 1/2}(\tau_1,\tau_2; m)/D_{0,\, 0}(\tau_1,\tau_2;
m)$ diverges when $\tau_1$ and $\tau_2$ vanish, we consider the
limits $\tau_1=0$ and $\tau_2\rightarrow 0$ for the integrand of
the free energy to see the apparently divergent behavior of
it~\cite{PandoZayas:2002hh,Greene:2002cd}. This limit is sufficient to
obtain the Hagedorn temperature, but it is subtle
  whether this limit captures the nature of the Hagedorn
temperature (see the results from Brower, Lowe and Tan
\cite{Brower:2002zx}). In the above limit,
$m$ goes to infinity which make it difficult to extract the
singular behavior of $D_{b_1,\, b_2}(\tau_1,\tau_2; m)$. To avoid
this difficulty, one can use the following modular
property~\cite{Takayanagi:2002pi,Bergman:2002hv}
\be \Big|D_{b_1,\, b_2}\Big(\tau_1,\tau_2 ; m \Big)\Big|^2 =
\Big|D_{b_2,\, -b_1} \Big(-\frac{\tau_1}{|\tau|^2},
\frac{\tau_2}{|\tau|^2} ; m|\tau|\Big)\Big|^2=\Big|D_{b_1,\,
b_2+b_1}\Big(\tau_1+1,\tau_2 ; m \Big)\Big|^2\,, \label{modular}
\ee
with $\tau =\tau_1 +i \tau_2$. In our cases $b_1,\, b_2=0,\, 1/2$,
it reduces to
\[ D_{b_1,\, b_2}\Big(\tau_1,\tau_2 ; m \Big) = D_{b_2,\, -b_1}
\Big(-\frac{\tau_1}{|\tau|^2}, \frac{\tau_2}{|\tau|^2} ;
m|\tau|\Big)=D_{b_1,\, b_2+b_1}\Big(\tau_1+1,\tau_2 ; m \Big)\,.
\]
%
% This modular
%transformation helps to extract the singular behavior of the
%integrand.
Under this transformation, the integrand in the free energy
expression becomes
%can be shown to be in the zero point energy, $\Delta_b(m)$,
%$(Z_{1/2,\, 0}(\tau_1,\tau_2 ; m/3)/Z_{0,\, 0}(\tau_1,\tau_2;
%m/3))^4 \times (Z_{1/2,\, 0}(\tau_1,\tau_2; m/6)/Z_{0,\,
%0}(\tau_1,\tau_2 ; m/6))^4$
%as
%
\bea  &&  \left[\frac{D_{0,\, 1/2}(\tau_1,\tau_2 ;
\frac{m}{3})}{D_{0,\, 0}(\tau_1,\tau_2; \frac{m}{3})}\right]^4
\left[\frac{D_{0,\, 1/2}(\tau_1,\tau_2; \frac{m}{6})}{D_{0,\,
0}(\tau_1,\tau_2 ; \frac{m}{6})}\right]^4
\nn \\   &&    \nn \\
&= & \left[\frac{D_{1/2,\, 0}(-\frac{\tau_1}{|\tau|^2},
\frac{\tau_2}{|\tau|^2} ; \frac{m}{3}|\tau|)}{D_{0,\,
0}(-\frac{\tau_1}{|\tau|^2},\frac{\tau_2}{|\tau|^2} ;
\frac{m}{3}|\tau|)} \right]^4  \left[\frac{D_{1/2,\, 0}
(-\frac{\tau_1}{|\tau|^2},\frac{\tau_2}{|\tau|^2} ;
\frac{m}{6}|\tau|)}{D_{0,\, 0} (-\frac{\tau_1}{|\tau|^2},
\frac{\tau_2}{|\tau|^2}  ; \frac{m}{6}|\tau|)} \right]^4~.
%\\ \!\!\!\!\!\!\!\!\!\! &&  \!\!\!\!\!
\eea
In the limit $\tau_1=0$, $\tau_2\rightarrow 0$, $m|\tau|=l\beta
\mu/(2\sqrt{2} \pi)$ is finite  and thus the leading term of the
integrand is readily expressed  as
\bea \exp\left[\frac{2\pi}{\tau_2}\bigg\{
4\Delta_{1/2}\Big({\textstyle\frac{l\beta\mu}{6\sqrt{2}\pi}}\Big)
-4\Delta_{0}\Big({\textstyle\frac{l\beta\mu}{6\sqrt{2}\pi}}\Big)+
4\Delta_{1/2}\Big({\textstyle\frac{l\beta\mu}{12\sqrt{2}\pi}}\Big)
-4\Delta_{0}\Big({\textstyle\frac{l\beta\mu}{12\sqrt{2}\pi}}\Big)\bigg\}
\right] \,, \nn\eea
%
%
%
% \exp\bigg[\frac{2}{\tau_2}
%\frac{r\beta\mu}{3\sqrt{2}\pi}\sum_{p=1}^{\infty}\frac{1-(-1)^p}{p}
%\bigg(2K_1 \Big(\frac{r\beta\mu}{3\sqrt{2}} p\Big) +  K_1
%\Big(\frac{n\beta\mu}{6\sqrt{2}} p\Big)\bigg)\bigg] \,,\eea
%
The asymptotics of the free energy, then, reads as
\bea \!\!\!\!\!\!\!\!\!\!  F &\sim& -\frac{L}{4\pi^2\a} \sum_{l=1,
{\rm odd}}^\infty \int^\infty_0
\frac{d\tau_2}{\tau^{~2}_2}{}~~{}\exp
\bigg[-\frac{l^2\beta^2}{4\pi\a\tau_2}\bigg]
 \nn \\\!\!\!\!\!\!\!\!\!\!   &&\nn \\ \!\!\!\!\!\!\!\!\!\!   &&{}~{}
\times \exp\bigg[ \frac{2\pi}{\tau_2} \bigg\{
4\Delta_{1/2}\Big({\textstyle\frac{l\beta\mu}{6\sqrt{2}\pi}}\Big)
-4\Delta_{0}\Big({\textstyle\frac{l\beta\mu}{6\sqrt{2}\pi}}\Big)+
4\Delta_{1/2}\Big({\textstyle\frac{l\beta\mu}{12\sqrt{2}\pi}}\Big)
-4\Delta_{0}
\Big({\textstyle\frac{l\beta\mu}{12\sqrt{2}\pi}}\Big)\bigg\}\bigg]\,.
\label{free2}\eea

The smallest temperature which gives rise to the singularity of
the above free energy comes from the $l=1$ case, which is taken as
the Hagedorn temperature $T_H = 1/\beta_H$:
\bea  \frac{\beta^2_H}{8\pi^2\a} &=&
4\left[\Delta_{1/2}\Big({\textstyle\frac{\beta_H\mu}{6\sqrt{2}\pi}}\Big)
-\Delta_{0}\Big({\textstyle\frac{\beta_H\mu}{6\sqrt{2}\pi}}\Big)\right]+
4\left[\Delta_{1/2}\Big({\textstyle\frac{\beta_H\mu}{12\sqrt{2}\pi}}\Big)
-\Delta_{0}\Big({\textstyle\frac{\beta_H\mu}{12\sqrt{2}\pi}}\Big)\right]
\label{Hagedorn} \\  && \nn \\ &=&
\frac{\beta_H\mu}{3\sqrt{2}\pi^2}\sum_{p=1}^\infty
\frac{1-(-1)^p}{p}
\bigg(2K_1\Big({\textstyle\frac{\beta_H\mu}{3\sqrt{2}} p}\Big) +
K_1\Big({\textstyle\frac{\beta_H\mu}{6\sqrt{2}} p}\Big) \bigg) \,.
\nn \eea
Hagedorn temperature is often defined as the temperature at which
the thermal winding mode of the string becomes
massless~\cite{Sathiapalan:1986db,Kogan:jd,atickwitten}. As will
be shown in the next section, this gives the same result. It is
interesting to observe that, if the effective tension of the
thermal winding string, depending on the temperature, is defined
as follows:
\[
\a_{{}_{eff}}(\beta\mu)
=\left(4\left[\Delta_{1/2}\Big({\textstyle\frac{\beta\mu}{6\sqrt{2}\pi}}\Big)
-\Delta_{0}\Big({\textstyle\frac{\beta\mu}{6\sqrt{2}\pi}}\Big)\right]+
4\left[\Delta_{1/2}\Big({\textstyle\frac{\beta\mu}{12\sqrt{2}\pi}}\Big)
-\Delta_{0}\Big({\textstyle\frac{\beta\mu}{12\sqrt{2}\pi}}\Big)\right]
\right)\a~,
\]
then, the leading term ($l=1$) in the free energy~(\ref{free2}),
after a suitable rescaling in $\tau_2$, can be written as
\bea F \sim -\frac{L}{4\pi^2\a_{{}_{eff}}} \int^\infty_0
\frac{d\tau_2}{\tau^{~2}_2}{}~~{}\exp
\bigg[-\frac{\beta^2}{4\pi\a_{{}_{eff}}\tau_2}+
\frac{2\pi}{\tau_2}\bigg] \,, \eea
and the Hagedorn temperature becomes
\[
\beta_H = 2\pi\sqrt{2\,\a_{{}_{eff}}(\beta_H\mu)}\,.
\]
Note that these expressions on the free energy and the Hagedorn
temperature take the same forms as those of the string theory on
the flat spacetime and reduce to the flat spacetime case smoothly
as $\mu$ goes to zero.

%In the limit, $\mu \rightarrow 0$ or $\beta \rightarrow 0$, the
%effective tension reduces to the

For the small value of $\mu$, the Hagedorn temperature can be
expressed  explicitly as a perturbative series in terms of
$\sqrt{\a}\mu$. Since the series expansion for the zero point
energy is given by~(see the Appendix for the full expression and
its derivation),
\be \Delta_{1/2}(x)-\Delta_{0}(x) = \frac{x}{\pi}\sum_{p=1}^\infty
\frac{1-(-1)^p}{p}K_1( 2\pi x p) = \frac{1}{8} -\frac{x}{2} +
x^2\ln 2 + {\cal O}(x^3)\,, \ee
the Hagedorn temperature up to $\mu^2$ order is
\be T_H = \frac{1}{2\pi
\sqrt{2\a}}\bigg(1+\frac{1}{2}{}\sqrt{\a}\mu +
\Big(\frac{1}{8}-\frac{5}{18}\ln 2\Big)\,
(\sqrt{\a}\mu)^2\bigg)\,. \ee

On the other hand, the large $\mu$ asymptotics of the Hagedorn
temperature can be obtained as
\be \frac{\beta^2_H}{32\pi^2\a} \sim
\frac{1}{\pi}\sqrt{\frac{\beta_H \mu}{12\sqrt{2}\pi}} {}~{}
\exp\bigg[-\frac{\beta_H \mu}{6\sqrt{2}}\bigg]
~~~\stackrel{\mu\rightarrow\infty}{\longrightarrow} 0 \quad ;
\quad T_H \stackrel{\mu\rightarrow\infty}{\longrightarrow}
\infty\,, \ee
through the asymptotic expansion of the zero point energy~:
\be \Delta_{1/2}(x)-\Delta_{0}(x)=\frac{x}{\pi}\sum_{p=1}^\infty
\frac{1-(-1)^p}{p}K_1(2\pi x p) \sim \frac{\sqrt{x}}{\pi} {}~{}
e^{-2\pi x}\,.\ee
The free energy in this limit becomes identical with the one in
the IIB case~\cite{semenoff}
\be F = -\frac{L\pi}{6\beta^2}\,, \label{2dfree}\ee
which shows a kind of the  universality  in the $\mu\rightarrow
\infty$ limit. Note that this is a typical free energy behavior of
two dimensional field theories at high
temperature~\cite{atickwitten}. It strongly indicates, at least in
the weakly coupled regime, that strings  on the pp-wave geometry,
in the large $\mu$ limit, effectively live on the two-dimensional
spacetime. The AdS/CFT correspondence suggests that the dual
theory of the M theory on maximally supersymmetric pp-wave would
be related to some common sector of six-dimensional (2,0) theory
or ${\cal N}=8$ three-dimensional conformal field theory. Since
our IIA pp-wave geometry comes from this eleven-dimensional
pp-wave, our IIA string theory should be deeply related to these
dual theories. It would be interesting to see the implications for
dual theories  of the behavior of free energy in the large $\mu$
limit.

%%%%%%%%%%%%%%%%%%%%%%%%%%%%%%%%%%%%%%%%%%%%%%%
\section{Path integral approach}

The path integral formalism is much better, compared to the
operator approach, in obtaining the genus-one free energy
expression with the manifest modular invariance. It also
incorporates the zero point energy more naturally and is
appropriate for extending to higher genera.

In this section, we use the path integral approach to obtain the
thermal one loop partition function and the Hagedorn temperature.
As will be shown in the next section, these agree perfectly with
the results given by the operator method. We follow the standard
formalism of thermal string theory and extend to GS superstring
theory on pp-wave. In the following subsections, we present all
the involved quantities in the path integral method and clarify
various issues including the modular invariance and the zero point
energy introduced in the operator method by hand.

\subsection{GS superstring at finite temperature}
In the finite temperature field theory, the time coordinate is
Wick-rotated and compactified with the circumference $\beta$.
Accordingly, the corresponding process in the finite temperature
string theory is to euclideanize the target space coordinate,
$X^0$, to  $X^0_{E}$ with the compactification radius $\beta$.
Along the compactified direction $X^0_{E}$, strings have the
winding modes, denoted by $r$, as well as the momentum modes,
denoted by $l$ and, thus, have the following periodicity
conditions :
\bea & X^{0}_{E}(\sigma^1+2\pi,\sigma^2) =
X^{0}_{E}(\sigma^1,\sigma^2)+r\beta\,,\nn  \\  & \label{bbc} \\  &
X^{0}_{E}(\sigma^1+2\pi\tau_1,\sigma^2+2\pi\tau_2) =
X^{0}_{E}(\sigma^1,\sigma^2)+l\beta\,, \nn \eea
where we also euclideanized worldsheet time as $\sigma^2=
i\sigma^0$. The classical configurations which satisfy these
boundary conditions are, up to constant term, given by
\be X_{wind}(\sigma^1,\sigma^2) = \frac{r\beta}{2\pi} \sigma^1 +
\frac{(l-r\tau_1)\beta}{2\pi\tau_2}\sigma^2 \,.
\label{thermwind}\ee
  Though the
spacetime Wick rotation on the pp-wave geometry itself
~\cite{Russo:2002rq} is problematic, we can still perform the Wick
rotation in the path integral of GS superstring on the pp-wave.

We start from the covariant IIA action~\cite{hyunandshin1} which
comes from the double dimensional reduction of the kappa gauge
fixed supermembrane actions on the eleven-dimensional pp-wave and
use the conformal gauge for the worldsheet metric. For
convenience, we perform the field redefinitions in the fermions as
\be  S^{a}_1=
          -\,(\gamma^4\,\psi^{1}_{-})^{a}\,,  \quad
          S^{a}_2 = \,\psi^{2~a}_{+}~~\,;\qquad
S'^{\,a}_1=\,(\gamma^4\,\psi^{1}_{+})^{a}\,, \quad
S'^{\,a}_2=\,\psi^{2~a}_{-} \,. \ee
Then the action becomes
\bea  S_E &=&  \frac{1}{8\pi\a}\int d z d\bar{z} \bigg[
-4(\partial X^+{\bar\partial} X^- + {\bar\partial} X^+\partial
X^-) + 4\partial X^I{\bar\partial}X^I -4 A(X^I)\partial
X^+\bar{\partial}X^+
  \nn \\ &&  \nn \\
&&{}~~~~~{} +8\,i \Big\{\partial X^+ ( S_1\bar{\partial} S_1 +
S'_1\bar{\partial}S'_1) + {\bar\partial} X^+(
      S_2\partial S_2 +  S'_2\partial S'_2)
     \nn \\ && \nn \\
&&\hskip2cm -({\textstyle\frac{\mu}{3}})\partial
X^+{\bar\partial}X^+ (2S_2 S_1 + S'_2S'_1)
      \Big\}\bigg]\,, \label{ELCA}
\eea
where $z = \sigma^1+i\sigma^2$ and ${\bar z}= \sigma^1-i\sigma^2$
with $dz d{\bar z} = 2d\sigma^1d\sigma^2$. For the calculation of
the one loop amplitude via the path integral approach in the
Green-Schwarz(GS) superstring, we should take into account the
boundary conditions for the GS fermions, $S^a$. They are spacetime
fermions and, thus, should be antiperiodic along the thermal
direction $X^0_{E}$ in the finite temperature formalism. Therefore
the boundary conditions for the fermions, consistent with the
periodicity conditions (\ref{bbc}), are~\cite{Carlip:1986cy}
\be S^a_1(z+2\pi ) = (-1)^{r}S^a_1(z)\,, \quad  S^a_1(z+2\pi\tau)
= (-1)^{l}S^a_1(z)\,, \label{FerBound}\ee
with the same conditions for $S'_1(z)$ and similar ones for
$S_2(\bar{z}), S'_2(\bar{z})$ under $\bar{z}$ periodicity.

The one loop free energy  can be represented as a path integral
over the gauge fixed action as
\bea - \beta F &=&  Z_{T^2}(\beta) = \int_{\cal F}\frac{d\tau_1
d\tau_2}{2\tau_2} \sum_{r,\, l \in\, {\bf Z}}^{} \int_{T^2/(r,\, l)} {\cal
D}X^+ {\cal D}X^- {\cal D}b {\cal D}c ~ {\cal
D}X^I {\cal D}S_{1,\,2}{\cal D}S'_{1,\,2} \nn \\ && \nn \\
&&\hskip3cm \times\, e^{-S_E-S_{gh}}\,(\det
\partial X^+)^{-4}\, (\det {\bar \partial} X^+)^{-4}\,,
\label{path-integral}\eea
where $(r,\, l)$ denotes the boundary conditions
Eq.~(\ref{FerBound}) for the fermions $S^a_{1,\,2},~
S'^{\,a}_{1,\,2}$ on the torus, and $S_{gh}$ is the same ghost
action as in the flat space case. ${\cal F}$ denotes the
fundamental region of the torus
\be {\cal F}:= \Big\{~~(\tau_1\,,\tau_2)~~; ~~~ -\frac{1}{2} <
\tau_1 \leq \frac{1}{2}\,, \qquad  \tau_2 > 0 \,, \qquad |\tau|
\geq 1 ~~~\Big\}\,. \ee
Note that the determinant in the above expression comes from the
$\kappa$ symmetry gauge fixing\footnote{It has been
argued~\cite{Carlip:1986cy} that it is not easy to incorporate
$\kappa$ symmetry in the path integral of GS superstring
rigorously. Nevertheless, the path integral defined as
(\ref{path-integral}) will be shown to be consistent with the
 operator formalism. }.

We perform, firstly, the integration over the longitudinal
directions $X^{\pm}$. As usual, we separate the fields into the
classical part and the quantum fluctuation one as
\be X^{\pm} (z,\,\bar{z}) = X^{\pm}_{cl}+ \delta X^{\pm}\,, \qquad
X^{\pm}_{cl} = x^{\pm}_0 +
{\textstyle\frac{-i}{\sqrt{2}}}X_{wind}\,.\ee
The classical thermal winding configuration, $X^{\pm}_{cl}$, leads
to the action value
\be S_E [X^{\pm}_{cl}] = \frac{\beta^2 |r\tau-l|^2}{4\pi\a
\tau_2}=: S_{\beta}(r,\, l)\,,  \ee
and the path integrals over $X^{\pm}$ are given by those of
$\delta X^{\pm}$ around this configuration.

The computation of the path integral for the one loop free energy
can be performed analogously to that of the one loop vacuum
amplitude in the GS superstrings on the flat
spacetime~\cite{Kallosh:wv} and on the IIB
pp-wave~\cite{Hammou:2002bf}. First, we integrate over $\delta
X^-$ which leads to the delta functional $\delta [\,
\partial {\bar\partial}\delta X^+\,]= \delta [\delta
X^+]/\det(\partial {\bar\partial})$. The integration over $\delta
X^+$ with this delta functional restricts $\delta X^+$ to be zero
in the integrand and, as usual, $1/\det(\partial {\bar\partial})$
term cancels out with the $bc$ ghost contribution except for the
zero mode, $x^{\pm}_0$, contribution. The path integrals over the
zero modes, $x^{\pm}_0$, give rise to $\beta L/(4\pi^2\a\tau_2)$
where $L$ is the (infinite) length of the longitudinal direction
used in the operator method given earlier. Therefore the path
integrals over $X^{\pm}$ and Faddeev-Popov ghosts
 around the classical
configurations~(\ref{thermwind}) give
\be \sum_{r,\, l \in\, {\bf Z}}^{} \frac{\beta
L}{4\pi^2\a\tau_2}{}~{} e^{-S_{\beta}(r,\, l)}~~\,, \ee
which is the same as the one given
in~\cite{Polchinski:rq,atickwitten,Polchinski:zf}. In addition,
$X^{\pm}$ fields in the residual integrand are replaced with
$X^{\pm}_{cl}$.

After  the field independent rescaling of fermions, $S_{1,\,2},\,
S'_{1,\,2}$, whose Jacobian cancels out the determinant factor
$\det (\bar{\partial} X^+_{cl})\det (\partial X^+_{cl})$, the one
loop free energy becomes the path integral over the transverse
fields with the Euclidean action:
\bea S'^E[X^I,\,S] &=&  \frac{1}{8\pi\a}\int d z d\bar{z} \bigg[
4\partial X^I{\bar\partial}X^I
+({\textstyle\frac{m'}{3}})^2(X^i)^2 +
({\textstyle\frac{m'}{6}})^2(X^{i'})^2
  \nn \\ &&  \nn \\
&&{}~~~{} + S_1\bar{\partial} S_1 + S'_1\bar{\partial}S'_1
-S_2\partial S_2 -  S'_2\partial S'_2
+({\textstyle\frac{m'}{3}})S_2 S_1 +
({\textstyle\frac{m'}{6}})S'_2S'_1
      \bigg]\,, \qquad
\eea
where
\be m' := \frac{\mu\beta}{2\sqrt{2}\pi\tau_2}|r\tau-l|\,.
\label{mprime}\ee
As a result, the one loop free energy for a closed string gas can
be written as
\be - \beta F =  Z_{T^2}(\beta) = \int_{\cal F}\frac{d\tau_1
d\tau_2}{2\tau_2} \sum_{r,\, l \in\, {\bf Z}}^{} \frac{\beta
L}{4\pi^2\a\tau_2}{}~{} e^{-S_{\beta}(r,\, l)}{}~{}
Z_{trans}^{(r,\, l)}(\tau_1,\tau_2) \,,\label{pathfree} \ee
where $Z_{trans}^{(r,\, l)}(\tau_1,\tau_2)$ denotes the path
integrals over the transverse fields:
\be Z_{trans}^{(r,\, l)}(\tau_1,\tau_2) := \int_{T^2/(r,\, l)}
{\cal D} X^I{\cal D}S^a_{1,\,2}{}~~{} e^{-S'^E[X^I,\,S]}\,.
\label{path}\ee
Note that $(r,\,l)=(0,0)$ part of the above free energy expression
corresponds to the one loop vacuum amplitude, which is zero in our
path integral formulation in parallel with the type IIB
case~\cite{Hammou:2002bf}. We will give more comments on this
vacuum amplitude later.

To get the transverse partition function~(\ref{path}),  it is
convenient~\cite{Alvarez:1985fw,Alvarez:1986sj} to use the new
coordinates, $\xi^1,\xi^2$, defined by $z=\xi^1 + \tau\xi^2,\,
\bar{z}=\xi^1 + {\bar\tau}\xi^2$, which give $dz d\bar{z} =
d^2\xi{}~{} 2\tau_2$ and
$\partial\bar{\partial}=(1/2\tau_2)^2[|\tau|^2\partial_1^2
+\partial_2^2 -2\tau_1\partial_1\partial_2]$. The various fields
$X^i,X^{i'},S^a_{1,\,2},S^{'\,a}_{1,\,2}$ on the torus are Fourier
expanded as
\bea  X(\xi_1,\xi_2) &=& \sum_{n_1,n_2}x_{n_1,n_2} {}~{}e^{
in_1\xi_1 +  i n_2\xi_2}\,,
 \\       &&              \nn     \\
S(\xi_1,\xi_2)&=& \sum_{n_1,n_2}S_{n_1,n_2}
\exp\bigg[ i\Big(n_1+{\textstyle\frac{1-(-1)^{r}}{4}}\Big)\xi_1 +
i \Big(n_2+{\textstyle\frac{1-(-1)^{l}}{4}}\Big)\xi_2\bigg]\,,
\eea
and the eigenvalues of those fields are given by
\bea  && (-4\partial{\bar\partial}+M^{\,2})X =
E_{n_1,n_2}X\,;\quad
E_{n_1,n_2}=\Big(\frac{1}{\tau_2}\Big)^2|n_1\tau-n_2|^2
+ M^{\,2}\,, \nn\\
&&\nn\\ &&(-4\partial{\bar\partial}+M^{\,2})S =
E^{(r,\,l)}_{n_1,n_2}S\,;\quad
E^{(r,\,l)}_{n_1,n_2}=\Big(\frac{1}{\tau_2}\Big)^2|
(n_1+{\textstyle\frac{1-(-1)^{r}}{4}})\tau
-(n_2+{\textstyle\frac{1-(-1)^{l}}{4}})|^2+M^{\,2}\,, \nn\eea
where $M$ denotes $\frac{m'}{3}$ or $\frac{m'}{6}$.
%\newline

\subsection{Complex scalar field with twisted boundary
condition}

The above type of path integrals was calculated in
\cite{Itzykson}. In this subsection we present the calculation in
our context and clarify several points. Since the generalization
to other cases are straightforward, it will be sufficient to
consider the path integral of a massive complex bosonic field
$\phi$ of mass $M$ with the twisted boundary conditions
$\phi(\xi_1+2\pi,\xi_2)=e^{2\pi ib_1}\phi(\xi_1,\xi_2)$ and
$\phi(\xi_1,\xi_2+2\pi)=e^{2\pi ib_2}\phi(\xi_1,\xi_2)$. The
energy eigenvalues with these boundary conditions are given by
\be E^{(b_1,b_2)}_{n_1,n_2}=\Big(\frac{1}{\tau_2}\Big)^2|
(n_1+b_1)\tau -(n_2+b_2)|^2+M^{\,2}\,,\ee
and the partition function becomes
\be Z_{\phi}(M):=\int {\cal D}\phi{\cal D}\bar{\phi} e^{-\int
d^2\xi \tau_2 \bar{\phi}(-4\partial{\bar\partial}+M^2)\phi} =
\prod_{n_1,n_2}(\tau_2E^{(b_1,b_2)}_{n_1,n_2})^{-1}\,, \ee
where we have absorbed the overall constant in the action by the
field redefinition of $\phi$. As usual, this determinant can be
regulated and evaluated using the $\zeta$-function method by
considering
\be G(s) := \sum_{n_1,n_2\in {\bf Z}}\left(
\frac{\mu_R^2}{\tau_2E^{(b_1,b_2)}_{n_1,n_2}}\right)^s\,, \qquad
{\rm Re}{}~{} s  > 1 \,,\label{zetaG}\ee
where we have introduced the renormalization scale $\mu_R$ to
insure the function $G(s)$ dimensionless in the $\zeta$-function
scheme. As is well-known, this corresponds to the introduction of
the cosmological constant term in the string worldsheet action as
a counter term. Note also that we have included the factor
$\tau_2$ which comes from the Lagrangian measure of the coordinate
transformation from $(z,\bar{z})$ to $(\xi^1,\xi^2)$. As was
pointed out in \cite{Itzykson}, it is required to perform the
ultraviolet renormalization in the above determinant
expression\footnote{In \cite{Itzykson}, ``specific heat'' is
treated as the quantity for the renormalization prescription.},
and our insertion of the renormalization scale, $\mu_R$, takes
care of that. By the analytic continuation of $s$, the
renormalized determinant can be evaluated through
\be e^{G^{\prime}(0)}=
\prod_{n_1,n_2}\frac{\mu^2_R}{\tau_2E^{(b_1,b_2)}_{n_1,n_2}}\,,
\qquad G^{\prime}(s) := \frac{d}{d s}G(s)\,.\ee
It is sufficient to consider $G(s)$ up to the linear order in $s$
to get $G'(0)$. Denoting $a = (n_1 + b_1)\tau_1 - b_2$ and $c =
\tau_2\sqrt{(n_1 + b_1)^2+M^2}$, we get
\bea G(s) &=& \sum_{n_1,n_2\in {\bf Z}}
\frac{(\mu^2_R\tau_2)^s}{[(n_2 + a)^2 + c^2]^s} \\ && \nn \\
&=& (\mu^2_R\tau_2)^s\sum_{n_1\in {\bf Z}}
\bigg[\sqrt{\pi}c^{1-2s}\frac{\Gamma(s-\half)}{\Gamma(s)} +
\frac{4\pi^s c^{\half-s}}{\Gamma(s)}
\sum_{p=1}^{\infty}\frac{\cos(2\pi a
p)}{p^{\half-s}}K_{\half-s}(2\pi c p)\bigg]\,,\nn\eea
where we have used the formula in the Appendix, (\ref{BesselK}),
for $n_2$ summation. Using the formula~(\ref{BesselK}) once again,
we can see that the first term in the sum can be expanded in terms
of $s$ as
\bea
\frac{\Gamma(s-\frac{1}{2})}{\Gamma(s)}\sum_{n_1}\frac{\sqrt{\pi}}{c^{2s-1}}
&=& \pi \frac{\Gamma(s-1)}{\Gamma(s)} \tau_2^{1-2s} M^{2-2s} +
\frac{4\pi^s \tau_2^{1-2s} M^{1-s}}{\Gamma(s)} \sum_{p=1}^{\infty}
\frac{\cos (2\pi b_1 p)}{p^{1-s}} K_{1-s}(2\pi M p)\,, \nn \\
 && \nn \\
 &=&  \pi\tau_2M^2+ s{}~{} \bigg[\pi\tau_2 M^2 \Big( -1 +
\ln M^2\tau_2^2\Big) - 4\pi\tau_2\Delta_{b_1}(M)\bigg]+ {\cal
O}(s^2)\,,\nn \eea
where $\Delta_b(m)$, the function defined by~(\ref{delta}),
appears naturally in this approach. The second term inside the sum
is also expanded as
\be   \frac{4\pi^s c^{\half-s}}{\Gamma(s)}
\sum_{p=1}^{\infty}\frac{\cos(2\pi a
p)}{p^{\half-s}}K_{\half-s}(2\pi c p) = s\,\bigg[ 4\, c^{\half}
\sum_{p=1}^{\infty}\frac{\cos(2\pi a p)}{p^{\half}}K_{\half}(2\pi
c p)\bigg] + {\cal O}(s^2)\,. \ee
To proceed, note that $K_{\half}(z) = \sqrt{\frac{\pi}{2z}}\,
e^{-z}$ and
\be 2 \sum_{p=1}^{\infty}\frac{\cos(2\pi a p)}{p}
        e^{-2 \pi c p}
= \sum_{p=1}^{\infty} \frac{1}{p}(q^p + \bar{q}^p) = -2|\ln(1-q)|
\,, \ee
where $q = e^{2\pi i a-2\pi c}= e^{ 2\pi i (n_1 + b_1)\tau_1 -
2\pi\tau_2
          \sqrt{(n_1+b_1)^2+M^2} - 2\pi i b_2 }$.
Finally,
 with $(\mu^2_R\tau_2)^s = 1 + s\,
\ln(\mu^2_R\tau_2)+{\cal O}(s^2)$,  we obtain
\bea G'(0) &=& - \pi \tau_2M^2  + \pi
       \tau_2 M^2\ln(\frac{M^2\tau_2}{\mu^2_R}) -
       4\pi\tau_2 \Delta_{b_1}(M)      \nn \\
       & &{}~~{}- 2\ln\bigg|\prod^{\infty}_{n_1=-\infty} \Big( 1 -
       e^{ 2\pi i (n_1+b_1) \tau_1 - 2\pi \tau_2
       \sqrt{(n_1+b_1)^2 + M^2} - 2\pi i b_2 } \Big) \bigg|\,.
\eea

In all, the partition function for the massive complex scalar
field $\phi$ is given by
\be
 Z_{\phi}(M) = \left[e^{2\pi\tau_2\Delta^R_{b_1}(M)}\bigg|\prod_{n_1}
\Big(1-e^{ 2\pi i (n_1+b_1) \tau_1 - 2\pi \tau_2
       \sqrt{(n_1+b_1)^2 + M^2} - 2\pi i b_2 }\Big)\bigg|\right]^{-2}\,,
       \ee
where
\be \Delta^R_{b_1}(M) = \Delta_{b_1}(M) - \quarter
M^2\Big[\ln\frac{M^2\tau_2}{\mu^2_R} -1\Big]\,.\ee
The above $\Delta^R_{b_1}(M)$ is the so-called zero point energy
(or Casimir energy) on the worldsheet. But we emphasize that this
is the renormalized quantity, not the physical one in the context
of the renormalizable quantum field theory~\cite{Collins:xc}. To
get the physical quantity we need to impose the renormalization
condition. We will take it by the criterion
\be \Delta_b^{ph}(M)\longrightarrow 0\,, \qquad  {\rm when}\quad
M\longrightarrow\infty\,, \label{rencond}\ee
which is the standard choice in the Casimir energy
literature~\cite{Bordag:2001qi} (see
also~\cite{Santangelo:2000fj,Milton:1999ge}). This consideration
finally gives us the physical quantity
$\Delta_{b_1}^{ph}(M)=\Delta_{b_1}(M)$, and it will be economical
to write the partition function only with the physical zero point
energy as
\be Z_{\phi}(M) =
\left[e^{2\pi\tau_2\Delta_{b_1}(M)}\bigg|\prod_{n_1} \Big(1-e^{
2\pi i (n_1+b_1) \tau_1 - 2\pi \tau_2
       \sqrt{(n_1+b_1)^2 + M^2} - 2\pi i b_2 }\Big)\bigg|\right]^{-2}\,\,.
\label{bosonZ}\ee

\subsection{Manifestly modular invariant one loop free energy}

As a special case of the path integrals in the previous
subsection, the partition function of the massive real scalar
field $X$ is given by
\be Z_{X}(M) = \left[e^{2\pi\tau_2\Delta_0(M)}\prod_n
\Big(1-e^{-2\pi\tau_2|\omega_n|+2\pi i \tau_1
n}\Big)\right]^{-1}\,. \label{bosonZZ}\ee
Although GS fermions are not worldsheet fermions, in the path
integral formalism they are treated as Grassmann variables only
with the different boundary conditions from those of worldsheet
fermions. Hence, with a change according to the nature as the
Grassmann variables, the direct application of the above result
for a complex boson to GS fermions, $S^a$, leads to
\bea Z_{S}^{(r,\,l)}(M) &=& e^{2\pi\tau_2\Delta_{b_1}(M)}\prod_n
\Big(1-e^{-2\pi\tau_2\sqrt{(n+b_1)^2+M^2}+2\pi i \tau_1 (n +
b_1)-2\pi i b_2 }\Big) \nn \\ && \nn \\ &\equiv&
D_{b_1,\,b_2}(\tau_1,\tau_2;M)~~~\,; \qquad b_1 =
\frac{1-(-1)^r}{4}\,, \quad b_2 = \frac{1-(-1)^l}{4}\,.
\label{fermionZ} \eea
This justifies the form of $D_{b_1,\,b_2}(\tau_1,\tau_2;M)$ with
$\Delta_{b_1}(m)$ introduced in a somewhat ``ad hoc'' way
in~(\ref{Dbb}). Moreover, resurrecting the full arguments of
$G(s)$ as $G_{b_1,\, b_2}(s;\, \tau_1,\tau_2,M)$, one can observe
the following modular property of $G(s)$ from the translational
symmetry by integer steps in the infinite double summation over
$n_1$ and $n_2$
\be G_{b_1,\, b_2}\Big(s;\, \tau_1,\tau_2,M\Big)= G_{b_2,\,
-b_1}\Big(s;\,
-{\textstyle\frac{\tau_1}{|\tau|^2}},{\textstyle\frac{\tau_2}{|\tau|^2}}
,M|\tau|\Big)= G_{b_1,\, b_2+b_1}\Big(s;\,
\tau_1+1,\tau_2,M\Big)\,. \ee
Note that we can always take $0\le b_{1,\,2} < 1$ by a suitable
translation. These properties are transferred to $G'(0)$ which
explains the modular property of $D_{b_1,\,b_2}(M)$
in~(\ref{modular}).

After all these, the transverse partition function is given by
\bea Z_{trans}^{(r,\,l)}(\tau_1,\tau_2) &=&
\Big[Z_{X}({\textstyle\frac{m'}{3}})\,
Z_{S}^{(r,\,l)}({\textstyle\frac{m'}{3}}) \Big]^4 \Big[
Z_{X}({\textstyle\frac{m'}{6}})\,Z_{S}^{(r,\,l)}
({\textstyle\frac{m'}{6})}
\Big]^4 \nn \\
&& \label{transR} \\
&=&\bigg[\frac{D_{b_1,\,b_2}(\tau_1,\tau_2;\frac{m'}{3})}
{D_{0,0}(\tau_1,\tau_2;\frac{m'}{3})}\bigg]^4
\bigg[\frac{D_{b_1,\,b_2}(\tau_1,\tau_2;\frac{m'}{6})}
{D_{0,0}(\tau_1,\tau_2;\frac{m'}{6})}\bigg]^4 \,. \nn\eea

After plugging (\ref{transR}) in (\ref{pathfree}), the final
expression of the one loop free energy in the path integral
approach is given by
\be  - F = \int_{\cal F}\frac{d\tau_1 d\tau_2}{2\tau_2}\!
\sum_{r,\, l\in{\bf Z}} \frac{L}{4\pi^2\a\tau_2}{}~{} e^{-S_\beta
(r,\, l)} \bigg[\frac{D_{b_1,\,b_2}(\tau_1,\tau_2;\frac{m'}{3})}
{D_{0,0}(\tau_1,\tau_2;\frac{m'}{3})}\bigg]^4
\bigg[\frac{D_{b_1,\,b_2}(\tau_1,\tau_2;\frac{m'}{6})}
{D_{0,0}(\tau_1,\tau_2;\frac{m'}{6})}\bigg]^4\,, \label{freeF}\ee
where  $m'$ and $b_{1,2}$ are defined in (\ref{mprime}) and
(\ref{fermionZ}), respectively.

The transverse partition function has the following modular
property, as can be seen from that of $G(s)$,
\be Z_{trans}^{(r,\,l)}(\tau_1,\tau_2) = Z_{trans}^{(l,\,-r)}
(-{\textstyle\frac{\tau_1}{|\tau|^2}},
{\textstyle\frac{\tau_2}{|\tau|^2}}) = Z_{trans}^{(r,\,l+r
)}(\tau_1+1,\tau_2)\,, \label{Zmod}\ee
from which the modular invariance of the one loop free energy
(\ref{freeF}) is manifest.

\subsection{The modular invariance of vacuum amplitude and
the flat space limit}

Now we would like to comment on the thermal GS superstrings on
flat space.  If we  take the correct $m$ ( or $M$) $\rightarrow 0$
limit, we should obtain the same free energy expression as the one
obtained in the RNS formalism~\cite{atickwitten}. The only
subtlety in taking the limit arises in the term $Z_X(M)$. While
$Z_{X}(M)\,Z_{S}^{(r,\,l)}(M)=1$ for even $r$ and $l$ and for
non-zero $M$, as can be seen from (\ref{bosonZZ}) and
(\ref{fermionZ}), $Z_{X}(0)\,Z_{S}^{(r,\,l)}(0)=0$ for even $r$
and $l$ due to the fermion zero modes. This shows that the flat
space limit is not smooth in the transverse partition function in
our formulation. Adopting the correct zero mass limit
$Z^{(r,\,l)}_S(0)=0,\,$ for even $r$ and $l$, $[Z_X(0)]^8=
V_8/(4\pi^2\a \tau_2)^4|\eta(\tau)|^{16} $ and
\be
 D_{0,\,1/2}(\tau_1,\tau_2;0)
     = \bigg|\frac{\theta_2 (\tau)}{\eta(\tau)}\bigg|\,,\quad
     D_{1/2,\,0}(\tau_1,\tau_2;0)
     = \bigg|\frac{\theta_4 (\tau)}{\eta(\tau)}\bigg|\,,\quad
 D_{1/2,\,1/2}(\tau_1,\tau_2;0) =
 \bigg|\frac{\theta_3 (\tau)}{\eta(\tau)}\bigg|\,,
 \ee
one can see that the one loop free energy becomes
\bea -F &=& L\,V_8 \int_{\cal F}\frac{d\tau_1
d\tau_2}{2\tau_2}\bigg( \frac{1}{4\pi^2\alpha'\tau_2} \bigg)^5
\bigg|\frac{1}{\eta(\tau)}\bigg|^{24} \bigg\{ \sum_{r: even \atop
l: odd}
e^{-S_{\beta}(r,\,l)} |\theta_2(\tau)|^8\nn \\ && \nn \\
&& {}~~~{}\hskip1.5cm
 + \sum_{r: odd\atop l: even}
e^{-S_{\beta}(r,\,l)} |\theta_4(\tau)|^8  + \sum_{r: odd\atop l:
odd} e^{-S_{\beta}(r,\,l)}| \theta_3(\tau)|^8 \bigg\}\,.\eea
Using the Jacobi identity $\theta_3^{\,4}(\tau) -
\theta_2^{\,4}(\tau) -\theta_4^{\,4}(\tau)=0$, this can be put
into the form given in~\cite{atickwitten}
\bea \!\!\!\!\!\!-\frac{F}{V} &=& \frac{1}{4}\int_{\cal
F}\frac{d\tau_1 d\tau_2}{2\tau_2} \bigg(
\frac{1}{4\pi^2\alpha'\tau_2} \bigg)^5
\bigg|\frac{1}{\eta(\tau)}\bigg|^{24}
 \sum_{r,\,l\in {\bf
Z}}e^{-S_{\beta}(r,\,l)} \bigg\{(\theta_2^{\,4}\bar{\theta}_2^{\,
4}+\theta_3^{\,4}\bar{\theta}_3^{\, 4} +
 \theta_4^{\,4}\bar{\theta}_4^{\, 4}) \nn \\ \!\!\!\!\!\!&& \nn \\
\!\!\!\!\!\! && {}~{} +{}~{} e^{\pi
i(r+l)}(\theta_2^{\,4}\bar{\theta}_4^{\, 4}+
\theta_4^{\,4}\bar{\theta}_2^{\, 4}) -  e^{\pi
i\,r}(\theta_2^{\,4}\bar{\theta}_3^{\, 4}+
\theta_3^{\,4}\bar{\theta}_2^{\, 4}) -  e^{\pi i\,
l}(\theta_3^{\,4}\bar{\theta}_4^{\, 4}+
\theta_4^{\,4}\bar{\theta}_3^{\, 4}) \bigg\}\,, \eea
where $V = L\,V_8 $.

Finally, let us discuss the $(r,\,l)=(0,\,0)$ part of the free
energy~(\ref{freeF}), which is nothing but the one loop vacuum
amplitude.  The modular invariance of this amplitude, which is one
of the consistency check of the given metric as a string theory
background, is guaranteed by construction in our path integral
approach. In fact, since $m'=0$ when $(r,\,l)=(0,\,0)$, the one
loop vacuum amplitude becomes zero through the correct $m'=0$
limit ({\it viz} flat space limit).  This is in agreement with the
fact that the one loop vacuum amplitude of the superstring on the
globally supersymmetric  background  should be zero by the
cancellation of the bosonic and fermionic vacuum energies.

\section{Equivalence between operator and path integral methods}

In this section we show the complete agreement between operator
and path integral approaches. Apparently, two approaches give the
different integration and summation regions in the free energy
expression. The free energy (\ref{freeF}) from the path integral
approach is written as the integral over the fundamental region of
the torus, ${\cal F}$, with the summations over two indices $r$
and $l$, while the one in the operator method is given by the
integral over the half strip region,
 \be E := \Big\{~~(\tau_1\,,\tau_2)~~; ~~~
-\frac{1}{2} \leq \tau_1 \leq \frac{1}{2}\,, \qquad  \tau_2 \geq 0
\,, ~~~\Big\}\,, \ee
 with the summation over $l$ only.
 We follow the arguments for the flat space case to show
 the
equivalence of these two forms of the free energy through the
resummation along with the change of the integration region.

In the case of type II superstrings on flat
space~\cite{O'Brien:pn,McClain:1986id}, the resummation with the
change of the integration region  can be summarized as
\be \sum^{r,\,l\in {\bf Z}}_{(r,\,l)\neq (0,\,0)}\int_{\cal
F}\frac{d\tau_1d\tau_2}{2\tau_2^2}{}~{}e^{-S_\beta (r,\,
l)}K_{(r,\,l)}(\tau_1,\tau_2) = ~\sum^{ l\in {\bf Z}}_{ l\neq
0}\int_{E}\frac{d\tau_1d\tau_2}{2\tau_2^2}{}~{}e^{-S_\beta (0,\,
l)}K_{(0,\,l)}(\tau_1,\tau_2)\,, \ee
where $K_{(r,\,l)}(\tau_1,\tau_2)$ is a function of modular
parameter with the modular property
\be K_{(r,\,l)}(\tau_1,\tau_2) = K_{(l,\,-r)}
(-{\textstyle\frac{\tau_1}{|\tau|^2}},
{\textstyle\frac{\tau_2}{|\tau|^2}}) = K_{(r,\,l+r
)}(\tau_1+1,\tau_2)\,. \label{Kmod}\ee

 Note that since the transverse partition
function $Z^{(r,\,l)}_{trans}(\tau_1,\tau_2)$ has the same modular
property, given in~(\ref{Zmod}), as $K_{(r,\,l)}(\tau_1,\tau_2)$,
the above resummation with the change of the integration region
can be applied to our case as well. Moreover, since
$(r,\,l)=(0,0)$ term has no contribution in the free energy
(\ref{freeF}), it does not have any effect in the resummation with
the change of the integration region.

After applying this resummation with the change of the integration
region on the  free energy (\ref{freeF}), and by noting that $m'$
in (\ref{mprime}), for $r=0$, becomes $m$ in (\ref{mmod}), the
free energy from the path integral approach (\ref{freeF}) can be
written as
\bea \!\!\!\!\!\!\!\!\!\!\!\!\!\!\! - F
 &=&
\int_{E}\frac{d\tau_1d\tau_2}{2\tau_2}\frac{L}{4\pi^2\a\tau_2}
\Bigg\{ \sum^{l\in 2{\bf Z}}_{l \neq 0} {}~{}e^{-\frac{\beta^2
l^2}{4\pi\a \tau_2}}{}~~{} +  \\
\!\!\!\!\!\!\!\!\!\!&& \nn \\
&& \hskip3.25cm \sum^{l \in 2{\bf Z}+1}_{l}{}~{}e^{-\frac{\beta^2
l^2}{4\pi\a \tau_2}}
\bigg[\frac{D_{0,\,1/2}(\tau_1,\tau_2;{\textstyle\frac{m}{3}})}
{D_{0,\,0}(\tau_1,\tau_2;{\textstyle\frac{m}{3}})}\bigg]^4
 \bigg[\frac{
D_{0,1/2}(\tau_1,\tau_2;{\textstyle\frac{m}{6}})}
{D_{0,0}(\tau_1,\tau_2;{\textstyle\frac{m}{6}})}\bigg]^4\Bigg\}\,,
\nn\eea
which is identical with the result from the operator method given
in Eq.~(\ref{Free}).

As we mentioned earlier, the alternative definition of the
Hagedorn temperature is the temperature where the thermal winding
modes become massless. In order to see this definition also gives
the same result, we perform the ${\cal S}$-modular transformation,
$$e^{-S_{\beta}(r,\,l)}Z_{trans}^{(r,\,l)}(\tau_1,\tau_2) =
e^{-S_{\beta}(l,\,-r)}
Z_{trans}^{(l,\,-r)}(-\frac{\tau_1}{|\tau|^2},\frac{\tau_2}{|\tau|^2}
)\,,$$ which exchanges the winding and momentum modes and do the
resummation with the change of integration region. After these
computations on the free energy (\ref{freeF}), it becomes
\bea \!\!\!\!\!\!\!\!\!\!\!\!\!\!\! - F
 &=&
\int_{E}\frac{d\tau_1d\tau_2}{2\tau_2}\frac{L}{4\pi^2\a\tau_2}
\Bigg\{ \sum^{r\in 2{\bf Z}}_{r \neq 0} {}~{}e^{-\frac{\beta^2
r^2}{4\pi\a \tau_2}}{}~~{} +  \\
\!\!\!\!\!\!\!\!\!\!&& \nn \\
&& \hskip0.7cm \sum^{r\in 2{\bf Z}+1}_{r}{}~{}e^{-\frac{\beta^2
r^2}{4\pi\a \tau_2}}\left[\frac{D_{1/2,\,
0}(-\frac{\tau_1}{|\tau|^2}, \frac{\tau_2}{|\tau|^2} ;
\frac{m}{3}|\tau|)}{D_{0,\,
0}(-\frac{\tau_1}{|\tau|^2},\frac{\tau_2}{|\tau|^2} ;
\frac{m}{3}|\tau|)} \right]^4  \left[\frac{D_{1/2,\, 0}
(-\frac{\tau_1}{|\tau|^2},\frac{\tau_2}{|\tau|^2} ;
\frac{m}{6}|\tau|)}{D_{0,\, 0} (-\frac{\tau_1}{|\tau|^2},
\frac{\tau_2}{|\tau|^2}  ; \frac{m}{6}|\tau|)} \right]^4\Bigg\}\,,
\nn\eea
whose asymptotics is given by (\ref{free2}). This shows that the
summation index $l$ in the asymptotic expression of one loop free
energy (\ref{free2}) can be interpreted as the index for winding
modes. The lowest winding mode becomes tachyonic at the Hagedorn
temperature~\cite{atickwitten,Sugawara:2002rs}.

We end this section with remarks clarifying subtle points
associated with these two approaches. First, we need the
renormalization prescription (like Eq.~(\ref{rencond})) in the
path integral approach while it is already done implicitly in the
operator method. The normal ordering prescription in the operator
approach is a kind of renormalization prescription for free
fields, but it is not incorporated in the path integral.

Second, the factor $\tau_2$ in Eq.~(\ref{zetaG}) was not
explicitly included in \cite{Itzykson}, in contrast to the string
theory
literatures~\cite{Alvarez:1985fw,Alvarez:1986sj,PandoZayas:2002hh}.
Nevertheless, we should have the same physical quantity because it
comes from the Lagrangian measure of the coordinate
transformation. This can be understood from the fact that the
physical quantity we will get after imposing the renormalization
condition is independent of this factor (and, of course, the
renormalization scale, $\mu_R$). This is also obvious from the
fact that we can take the new renormalization scale as
$\tilde{\mu} = \mu_R/\sqrt{\tau_2}$ in Eq.~(\ref{zetaG}).

Third, the $\zeta$-function method gives us the regularized form
of the zero point energy directly, and this gives the identical
result with the one from the standard method for the zero point
(or Casimir) energy
\be E^{Casimir}_b(m) = \half
\sum_{n=-\infty}^{\infty}\sqrt{(n+b)^2+m^2}- \half
\int^{\infty}_{-\infty} d k \sqrt{(k+b)^2+m^2}\,,\ee
which satisfy the criterion $E^{Casimir}_b(m)\rightarrow 0$ when
$m\rightarrow \infty$. Though this $E^{Casimir}_b(m)$ is identical
with $\Delta_b(m)$, we do not need this regularization as a
separate treatment in the operator method or the path integral one
because we have already fixed the renormalization prescription in
the former one by the normal ordering and in the latter one by the
$\zeta$-function with a suitable renormalization condition.

Finally, since the Hagedorn temperature dependence on the RR flux
is entirely given by the difference of bosons and fermions zero
point energy as indicated in Eq.~(\ref{Hagedorn}), it does not
depend on whether we use the renormalized zero point energy or the
physical one\footnote{The previous numerical
mismatch~\cite{PandoZayas:2002hh,semenoff} for the Hagedorn
temperature in IIB strings on pp-wave comes from the incorrect
series expansion for the zero point
energy~\cite{PandoZayas:2002hh} and it has no relation with the
formalism or the renormalization.}.

\section*{Acknowledgments}
One of us (J.-D. P.) would like to thank the Yonsei Visiting
Research Center (YVRC) for its hospitality, where this work has
been completed. The work of S.H. was supported by Korea Research
Foundation Grant (KRF-2002-042-C00010).

%\newpage

\appendix

\begin{center}
\large{\textbf{Appendix}}
\end{center}
\setcounter{equation}{0}
\renewcommand{\theequation}{A.\arabic{equation}}

%%%%%%%%%%%%%%%%%%%%%%%%%%%%%%%%%%%%%%%%%%%%%%%%%

\section{Heuristic derivation of the one loop free energy}
In this appendix, we use the Coleman-Weinberg
formula~\cite{Polchinski:rq,Polchinski:zf} to obtain the one loop
free energy, which results in
\be - \beta F =  Z_{T^2}(\beta) = -i\int_{\cal F}\frac{d\tau_1
d\tau_2}{2\tau_2} \sum_{r,\, l \in\, {\bf Z}}^{} \frac{\beta
L}{(2\pi)^2}\int d p^{+} dp^{-}{}~{} e^{2\pi\tau_2\a
p^{+}p^{-}}{}~{} e^{-S_{\beta}(r,\, l)}{}~{} Z_{trans}^{(r,\,
l)}(\tau_1,\tau_2)\,.\label{freeCW} \ee
Recalling the Wick rotation $p^{0}=ip^{0}_{E}$ and $p^{\pm}
=(p^{0}\pm p^{9})/\sqrt{2}$, one can see that this is identical
with the previous one loop free energy~(\ref{pathfree}) through
\be \frac{\beta L}{(2\pi)^2}\int d p^{+} dp^{-}{}~{}
e^{2\pi\tau_2\a p^{+}p^{-}} = i \frac{\beta L}{4\pi^2\a\tau_2} \,.
\label{factorL}\ee
The transverse partition function, in this approach, is
represented as an operator trace~\cite{Sugawara:2002rs}
\be Z_{trans}^{(r,\, l)}(\tau_1,\tau_2)  \equiv
 \tr_{trans}\Big[ (-1)^{(l+1){\bf F}} e^{-2\pi\tau_2
 {\cal H}(r)+2\pi i \tau_1 {\cal P}}\Big]\,,\ee
where $ {\cal H}(r)$ is defined by the same form as~(\ref{lcham})
while the integer $n$ in $\omega_n$ and $\omega'_n$ is replaced by
$n+\frac{1-(-1)^r}{4}$.

The naive application to $(r,\,l)=(0,0)$ mode of the free energy
expression~(\ref{freeCW}) with the above operator form of the
transverse partition function gives
\bea - \beta F &=&  Z_{T^2}(\beta) \\ && \nn \\
&=& \int_{\cal F}\frac{d\tau_1 d\tau_2}{2\tau_2} \frac{-i\beta
L}{(2\pi)^2}\int d p^{+} d p^{-}{}~{}e^{2\pi\tau_2\a
p^{+}p^{-}}{}~~{}
 \tr_{trans}\Big[ (-1)^{{\bf F}} e^{-2\pi\tau_2
 {\cal H}+2\pi i \tau_1 {\cal P}}\Big]\,.\nn \eea
After $\beta$ is regarded as another longitudinal length, this is
identical with the one loop vacuum amplitude
in~\cite{Takayanagi:2002pi} except for the details in our IIA
setting. Apparently, it seems to be divergent, but, we believe, it
is the artifact of the light-cone gauge $X^{+}\sim \sigma^{0}$
chosen in the operator formalism. As we showed in the main text,
the same quantity without the light-cone gauge fixing in the path
integral formalism is zero. Since our background is consistent and
globally supersymmetric, the one loop vacuum amplitude should be
zero by the cancellation of the bosonic and fermionic vacuum
energies. Therefore the above expression of one loop free energy
should be understood as the one without $(r,\,l)=(0,0)$ term in
the summation.

%%%%%%%%%%%%%%%%%%%%%%%%%%%%%%%%%%%%%%%%%%%%%%%%%

\setcounter{equation}{0}
\renewcommand{\theequation}{B.\arabic{equation}}

\section{Useful formulae}

In this section, we derive, following the methods in
Ref.~\cite{Ambjorn:1981xw}, the series expansion formula of the
function, $\Delta_b(m)$, defined by Eq.~({\ref{delta}) in the more
general setting than that in Ref.~\cite{semenoff}. Since this
expansion can cover the more general twisted boundary conditions,
it may be useful for studying the small $\mu$ dependence of the
Hagedorn temperature in orbifolded or DLCQ pp-wave case.

Firstly, let us consider the following function
\be {\cal S}(s)_{(b,m)} :=
\sum_{n=-\infty}^{\infty}\frac{1}{[(n+b)^2+m^2]^s}\,, \qquad {\rm
Re}(s) > \frac{1}{2} \,. \label{Sfunction}\ee
Note that we can take $0\le b <1$ without loss of the generality
from the translational symmetry. This function can be represented
through the Schwinger's proper time parametrization, $1/z^s
=\int^{\infty}_0 dt\, t^{s-1}\, e^{-tz}/\Gamma(s)$, as
\bea {\cal S}(s)_{(b,m)} &=&
\frac{1}{\Gamma(s)}\int_0^{\infty}dt{}~{}t^{s-1}{}~{}
e^{-tm^2}\sum_{n=-\infty}^{\infty}e^{-t(n+b)^2} \nn \\ &&\nn \\
&=&\frac{\sqrt{\pi}}{\Gamma(s)} \left[
\int_0^{\infty}dt{}~{}t^{(s-\half)-1}{}~{} e^{-tm^2} +
2\sum_{p=1}^{\infty}\cos(2\pi b p)
\int_0^{\infty}dt{}~{}t^{(s-\half)-1}{}~{}
e^{-tm^2-\frac{\pi^2p^2}{t}}\right]
\nn \\ && \nn \\
&=&\sqrt{\pi}m^{1-2s}\frac{\Gamma(s-\half)}{\Gamma(s)} +
\frac{4\pi^sm^{\half-s}}{\Gamma(s)}
\sum_{p=1}^{\infty}\frac{\cos(2\pi b
p)}{p^{\half-s}}K_{\half-s}(2\pi m p)\,, \label{BesselK}\eea
where we have used the Poisson resummation formula in the second
line and the integral representation of the modified Bessel
function $K_\nu(z)$ for the second term in the third line
\bea && \sum_{n=-\infty}^{\infty}e^{-t(n+b)^2} =
\sqrt{\frac{\pi}{t}}
\sum_{p=-\infty}^{\infty}e^{-\frac{\pi^2}{t}p^2+2\pi i b p} =
\sqrt{\frac{\pi}{t}}\Big[1+2\sum_{p=1}^{\infty}
e^{-\frac{\pi^2}{t}p^2}\cos(2\pi
bp)\Big]\,, \nn \\ && \nn \\
 &&{}~~{} K_{\nu}(z) = \half
{\textstyle(\frac{z}{2})}^{\nu}\int_0^{\infty}dt{}~{}t^{-\nu-1}
{}~{}e^{-t-\frac{z^2}{4t}}\,. \nn \eea

Next, note that ${\cal S}(s)_{(b,m)}$ can also be expanded as a
power series in terms of $m$ as,
\bea {\cal S}(s)_{(b,m)} & = &\left\{\ba{ll} \frac{1}{m^{2s}} +
2{\displaystyle\sum_{n=1}^{\infty}}
\frac{1}{n^{2s}}[1+\frac{m^2}{n^2}]^{-s}\,, & {}~~~{}b=0\,,
\\ & \nn \\
{\displaystyle\sum_{n=0}^{\infty}}\frac{1}{(n+b)^{2s}}
\Big[1+\frac{m^2}{(n+b)^{2}}\Big]^{-s}+
{\displaystyle\sum_{n=0}^{\infty}} \frac{1}{(n+1-b)^{2s}}
\Big[1+\frac{m^2}{(n+1-b)^{2}}\Big]^{-s}\,, & {}~~~{} 0 < b < 1\,,
\ea\right.
  \\ &&\nn \\
&=& \left\{\ba{ll}\frac{1}{m^{2s}}+2\zeta(2s)-2s\zeta(2s+2)m^2 +
2{\displaystyle\sum_{k=2}^{\infty}}
\frac{(-1)^k\Gamma(k+s)}{\Gamma(s)\Gamma(k+1)}\zeta(2k+2s)m^{2k}
\,, & {}~~~{} b=0\,, \\ & \\
\zeta(2s,b)+\zeta(2s,1-b)-s\Big[\zeta(2s+2,b)+\zeta(2s+2,1-b)\Big]m^2
&  \\ {}~~~~~{}+ {\displaystyle\sum_{k=2}^{\infty}}
\frac{(-1)^k\Gamma(k+s)}{\Gamma(s)\Gamma(k+1)}
\Big[\zeta(2k+2s,a)+\zeta(2k+2s,1-a)\Big] m^{2k} \,,  & 0<b<1\,,
\ea\right. \label{Series}\eea
where $\zeta(s)$ and $\zeta(s,q)$ are the (Riemann) zeta function
and the generalized (or Riemann-Hurwitz) zeta function,
respectively, defined as
\be \zeta(s) := \sum_{n=1}^{\infty}\frac{1}{n^s}\,,\qquad
\zeta(s,q) := \sum_{n=0}^{\infty}\frac{1}{(n+q)^s}\,,\quad q\neq
0,-1,-2,\cdots~. \ee
Some useful formulae for them are
\bea & \!\!\!\!\!\! \zeta(s)=\zeta(s,1)\,, & \qquad \zeta(s,\half)
 =(2^s-1)\zeta(s)\,, \nn \\ &  &\label{zetapro} \\
&{}~~~~~{} \zeta(-1,q) = \frac{1}{24}-\frac{1}{8}(2q-1)^2\,, &
\qquad \zeta(s,q) = \frac{1}{s-1} -\psi(q) + {\cal O}(s-1)\,,
 \nn\eea
where $\psi(q):=\frac{d}{dq}\ln\Gamma(q)$ with
$\psi(1)=-\gamma_E=-0.5772 ...$ and $\psi(1/2)=-\gamma_E-2\ln2$.
Here, $\gamma_E$ is the Euler constant.

 Now, we identify Eq.(\ref{BesselK}) with
Eq.(\ref{Series}) for sufficiently large and positive $s$-values,
and perform the analytic continuation to the entire $s$-complex
plane. Then, we can see that the function, $\Delta_b(m)$, can be
represented as
\be 2\Delta_b(m)\equiv -\frac{2m}{\pi}
\sum_{p=1}^{\infty}\frac{\cos(2\pi b p)}{p}K_{1}(2\pi m p) =
\lim_{s\rightarrow -\half}^{}\left[{\cal
S}(s)_{(b,m)}-\sqrt{\pi}m^{1-2s}
\frac{\Gamma(s-\half)}{\Gamma(s)}\right]\,. \label{ancon}\ee
When we take $s\rightarrow -\half$ in the above, the would-be pole
part $1/(s+\half)$ in the series form of ${\cal S}(s)_{(b,m)}$,
which comes from $\zeta(2s+2)$ or $\zeta(2s+2,b)+\zeta(2s+2,1-b)$
at $m^2$ order in Eq.~(\ref{Series}) , cancels out exactly with
the one from the Laurent expansion
\be \sqrt{\pi} m^{1-2s}\frac{\Gamma(s-\half)}{\Gamma(s)} = \half
m^2\Big[\frac{1}{s+\half} - \ln \frac{m^2}{4}+1\Big] +{\cal
O}(s+\half)\,. \ee
We emphasize that the above procedure through Eq.~(\ref{ancon})
does not involve the divergent expression which needs a
regularization but just gives us a mathematical relation by the
analytic continuation\footnote{Note that Eq.~(\ref{Sfunction}) is
used just as an intermediate step rather than the aim contrary to
Ref.~\cite{Ambjorn:1981xw}. We might, of course, take this
directly as a regularized definition for the zero point energy
like in Ref.~\cite{Ambjorn:1981xw}. But this requires the physical
consideration leading the renormalization condition. Therefore,
our logic is different from Ref.~\cite{Ambjorn:1981xw}~.}.

With Eq.s~(\ref{Series}),(\ref{zetapro}), we can finally get the
power series expansion for $\Delta_b(m)$ in terms of $m$ as,
\bea \Delta_b(m)
 = \left\{\ba{ll}-\frac{1}{12}+\half m+\quarter m^2
 \Big[\ln\frac{~m^2}{4}+2\gamma_E-1\Big]
 +{\displaystyle\sum_{k=2}^{\infty}}
 \frac{(-1)^k\Gamma(k-\half)}{\Gamma(-\half)\Gamma(k+1)}\zeta(2k-1)
 m^{2k} \,, & {}~~{} b = 0~,
 \\ & \!\!\! \\
\frac{1}{24}-\frac{1}{8}(2b-1)^2 + \quarter m^2
 \Big[\ln\frac{~m^2}{4}-\psi(b)-\psi(1-b)-1\Big] & \\ {}~~~{}
 + {\displaystyle \sum_{k=2}^\infty}
 \frac{(-1)^k \Gamma(k-\half)} {\Gamma(-\half)\Gamma(k+1)}
 \half\Big[\zeta(2k-1,b)+\zeta(2k-1,1-b)\Big] m^{2k} \,, &
 \!\!\!\!\!\! 0 < b <1~.
 \ea\right.
\eea
The special cases  $b=0$ and $b=1/2$ give us
\be \Delta_{1/2}(m) - \Delta_{0}(m) =
{\textstyle\frac{1}{8}}-\half m + m^2\ln 2 + {\displaystyle
\sum_{k=2}^\infty}
 \frac{(-1)^k \Gamma(k-\half)} {\Gamma(-\half)\Gamma(k+1)}
 (2^{2k-1}-2)\zeta(2k-1) m^{2k}\,, \label{diff}
\ee
which is the identical expression with the one in
Ref.~\cite{semenoff,Itzykson}.

\setcounter{equation}{0}
\renewcommand{\theequation}{B.\arabic{equation}}


\newpage


\begin{thebibliography}{99}



\bibitem{Polchinski:rq}
J.~Polchinski, ``String Theory. Vol. 1, 2'', Cambridge, Uk: Univ.
Pr. (1998).
%\href{http://www.slac.stanford.edu/spires/find/hep/www?irn=4634799}{SPIRES entry}



\bibitem{atickwitten}
J.~J.~Atick and E.~Witten,
%``The Hagedorn Transition And The Number Of Degrees Of Freedom Of String Theory,''
Nucl.\ Phys.\ B {\bf 310} (1988) 291.
%%CITATION = NUPHA,B310,291;%%


\bibitem{Hagedorn:st}
R.~Hagedorn,
%``Statistical Thermodynamics Of Strong Interactions At High-Energies,''
Nuovo Cim.\ Suppl.\  {\bf 3} (1965) 147.
%%CITATION = NUCUA,3,147;%%

\bibitem{huang}
K. Huang and S. Weinberg, Phys. Rev. Lett. {\bf 25}, 895 (1970).


\bibitem{PandoZayas:2002hh}
L.~A.~Pando Zayas and D.~Vaman,
%``Strings in RR plane wave background at finite temperature,''
[arXiv:hep-th/0208066].
%%CITATION = HEP-TH 0208066;%%


%\cite{Greene:2002cd}
\bibitem{Greene:2002cd}
B.~R.~Greene, K.~Schalm and G.~Shiu,
%``On the Hagedorn behaviour of pp-wave strings and N = 4 SYM theory at  finite R-charge density,''
Nucl.\ Phys.\ B {\bf 652} (2003) 105 [arXiv:hep-th/0208163].
%%CITATION = HEP-TH 0208163;%%

\bibitem{Sugawara:2002rs}
Y.~Sugawara,
%``Thermal amplitudes in DLCQ superstrings on pp-waves,''
Nucl.\ Phys.\ B {\bf 650} (2003) 75 [arXiv:hep-th/0209145].
%%CITATION = HEP-TH 0209145;%%


%\cite{Brower:2002zx}
\bibitem{Brower:2002zx}
R.~C.~Brower, D.~A.~Lowe and C.~I.~Tan,
%``Hagedorn transition for strings on pp-waves and tori with chemical  potentials,''
Nucl.\ Phys.\ B {\bf 652} (2003) 127 [arXiv:hep-th/0211201].
%%CITATION = HEP-TH 0211201;%%





\bibitem{Sugawara:2003qc}
Y.~Sugawara,
%``Thermal partition function of superstring on compactified pp-wave,''
[arXiv:hep-th/0301035].
%%CITATION = HEP-TH 0301035;%%




\bibitem{semenoff}
G.~Grignani, M.~Orselli, G.~W.~Semenoff and D.~Trancanelli,
%``The superstring Hagedorn temperature in a pp-wave background,''
[arXiv:hep-th/0301186].
%%CITATION = HEP-TH 0301186;%%

\bibitem{bla242} M. Blau, J. Figueroa-O'Farrill, C. Hall and G.
Papadopoulus,
%``A new maximally supersymmetric background of IIB
%superstring theory,''
JHEP {\bf 0201} (2001) 047, [arXiv:hep-th/0110242].






\bibitem{Metsaev:2001bj}
R.~R.~Metsaev,
%``Type IIB Green-Schwarz superstring in plane wave Ramond-Ramond  background,''
Nucl.\ Phys.\ B {\bf 625} (2002) 70 [arXiv:hep-th/0112044];
%%CITATION = HEP-TH 0112044;%%
R.~R.~Metsaev and A.~A.~Tseytlin,
%``Exactly solvable model of superstring in plane wave Ramond-Ramond  background,''
Phys.\ Rev.\ D {\bf 65} (2002) 126004 [arXiv:hep-th/0202109].
%%CITATION = HEP-TH 0202109;%%



\bibitem{Berenstein:2002jq}
D.~Berenstein, J.~M.~Maldacena and H.~Nastase,
%``Strings in flat space and pp waves from N = 4 super Yang Mills,''
JHEP {\bf 0204} (2002) 013 [arXiv:hep-th/0202021].
%%CITATION = HEP-TH 0202021;%%




\bibitem{hyunandshin1}
S.~Hyun and H.~Shin,
%``N = (4,4) type IIA string theory on pp-wave background,''
JHEP {\bf 0210} (2002) 070 [arXiv:hep-th/0208074].
%%CITATION = HEP-TH 0208074;%%

\bibitem{fig308}
J. Figueroa-O'Farrill, G. Papadopoulos,
%``Homogeneous Fluxes,
%Branes and a Maximally Supersymmetric Solution of M-theory,''
JHEP {\bf 0108} (2001)  036, [arXiv:hep-th/0105308].



%\cite{Frautschi:1971ij}
\bibitem{Frautschi:1971ij}
S.~Frautschi,
%``Statistical Bootstrap Model Of Hadrons,''
Phys.\ Rev.\ D {\bf 3} (1971) 2821.
%%CITATION = PHRVA,D3,2821;%%

%\cite{Carlitz:1972uf}
\bibitem{Carlitz:1972uf}
R.~D.~Carlitz,
%``Hadronic Matter At High Density,''
Phys.\ Rev.\ D {\bf 5} (1972) 3231.
%%CITATION = PHRVA,D5,3231;%%





%\cite{Bowick:az}
\bibitem{Bowick:az}
M.~J.~Bowick and L.~C.~Wijewardhana,
%``Superstrings At High Temperature,''
Phys.\ Rev.\ Lett.\  {\bf 54} (1985) 2485.
%%CITATION = PRLTA,54,2485;%%

%\cite{Bowick:1989us}
\bibitem{Bowick:1989us}
M.~J.~Bowick and S.~B.~Giddings,
%``High Temperature Strings,''
Nucl.\ Phys.\ B {\bf 325} (1989) 631.
%%CITATION = NUPHA,B325,631;%%

%\cite{Deo:1988jj}
\bibitem{Deo:1988jj}
N.~Deo, S.~Jain and C.~I.~Tan,
%``Strings At High-Energy Densities And Complex Temperature,''
Phys.\ Lett.\ B {\bf 220} (1989) 125;
%%CITATION = PHLTA,B220,125;%
N.~Deo, S.~Jain and C.~I.~Tan,
%``String Statistical Mechanics Above Hagedorn Energy Density,''
Phys.\ Rev.\ D {\bf 40} (1989) 2626;
%%CITATION = PHRVA,D40,2626;%
N.~Deo, S.~Jain, O.~Narayan and C.~I.~Tan,
%``The Effect of topology on the thermodynamic limit for a string gas,''
Phys.\ Rev.\ D {\bf 45} (1992) 3641.
%%CITATION = PHRVA,D45,3641;%%



%\cite{Hyun:2002wp}
\bibitem{hyunandshin2}
S.~Hyun and H.~Shin,
%``Solvable N = (4,4) type IIa string theory in plane-wave background and  D-branes,''
Nucl.\ Phys.\ B {\bf 654} (2003) 114 [arXiv:hep-th/0210158].
%%CITATION = HEP-TH 0210158;%%







\bibitem{Takayanagi:2002pi}
T.~Takayanagi,
%``Modular invariance of strings on pp-waves with RR-flux,''
JHEP {\bf 0212} (2002) 022 [arXiv:hep-th/0206010].
%%CITATION = HEP-TH 0206010;%%

%\cite{Bergman:2002hv}
\bibitem{Bergman:2002hv}
O.~Bergman, M.~R.~Gaberdiel and M.~B.~Green,
%``D-brane interactions in type IIB plane-wave background,''
JHEP {\bf 0303} (2003) 002 [arXiv:hep-th/0205183].
%%CITATION = HEP-TH 0205183;%%




%\cite{Sathiapalan:1986db}
\bibitem{Sathiapalan:1986db}
B.~Sathiapalan,
%``Vortices On The String World Sheet And Constraints On Toral Compactification,''
Phys.\ Rev.\ D {\bf 35} (1987) 3277.
%%CITATION = PHRVA,D35,3277;%%




%\cite{Kogan:jd}
\bibitem{Kogan:jd}
Y.~I.~Kogan,
%``Vortices On The World Sheet And String's Critical Dynamics,''
JETP Lett.\  {\bf 45} (1987) 709 [Pisma Zh.\ Eksp.\ Teor.\ Fiz.\
{\bf 45} (1987) 556].
%%CITATION = JTPLA,45,709;%%


%\cite{Russo:2002rq}
\bibitem{Russo:2002rq}
J.~G.~Russo and A.~A.~Tseytlin,
%``On solvable models of type IIB superstring in NS-NS and R-R plane wave  backgrounds,''
JHEP {\bf 0204} (2002) 021 [arXiv:hep-th/0202179].
%%CITATION = HEP-TH 0202179;%%




\bibitem{Carlip:1986cy}
S.~Carlip,
%``Loop Calculations For The Green-Schwarz Superstring,''
Phys.\ Lett.\ B {\bf 186} (1987) 141;
%%CITATION = PHLTA,B186,141;%%
%\bibitem{Carlip:1986cz} S.~Carlip,
%``Heterotic String Path Integrals With The Green-Schwarz Covariant Action,''
Nucl.\ Phys.\ B {\bf 284} (1987) 365.
%%CITATION = NUPHA,B284,365;%%





\bibitem{Kallosh:wv}
R.~Kallosh and A.~Y.~Morozov,
%``Green-Schwarz Action And Loop Calculations For Superstring,''
Int.\ J.\ Mod.\ Phys.\ A {\bf 3} (1988) 1943 [Sov.\ Phys.\ JETP
{\bf 67} (1988\ ZETFA,94,42-56.1988) 1540].
%%CITATION = IMPAE,A3,1943;%%






\bibitem{Hammou:2002bf}
A.~B.~Hammou,
%``One loop partition function in plane waves R-R background,''
JHEP {\bf 0211} (2002) 028 [arXiv:hep-th/0209265].
%%CITATION = HEP-TH 0209265;%%



%\cite{Polchinski:zf}
\bibitem{Polchinski:zf}
J.~Polchinski,
%``Evaluation Of The One Loop String Path Integral,''
Commun.\ Math.\ Phys.\  {\bf 104} (1986) 37.
%%CITATION = CMPHA,104,37;%%




%\cite{Alvarez:1985fw}
\bibitem{Alvarez:1985fw}
E.~Alvarez,
%``Strings At Finite Temperature,''
Nucl.\ Phys.\ B {\bf 269} (1986) 596.
%%CITATION = NUPHA,B269,596;%%

%\cite{Alvarez:1986sj}
\bibitem{Alvarez:1986sj}
E.~Alvarez and M.~A.~Osorio,
%``Superstrings At Finite Temperature,''
Phys.\ Rev.\ D {\bf 36} (1987) 1175.
%%CITATION = PHRVA,D36,1175;%%



\bibitem{Itzykson}
H. Saleur and C. Itzykson, ``Two-Dimensional Field Theories Close
to Criticality'', J. Statist. Phys. {\bf 48} (1987) 449.




\bibitem{Collins:xc}
J.~C.~Collins, ``Renormalization'', Cambridge, Uk: Univ. Pr.
(1984) 380p.
%``Renormalization. An Introduction To Renormalization,
%The Renormalization Group, And The Operator Product Expansion,''
%\href{http://www.slac.stanford.edu/spires/find/hep/www?irn=1341391}
%{SPIRES entry}





%\cite{Bordag:2001qi}
\bibitem{Bordag:2001qi}
M.~Bordag, U.~Mohideen and V.~M.~Mostepanenko,
%``New developments in the Casimir effect,''
Phys.\ Rept.\  {\bf 353} (2001) 1 [arXiv:quant-ph/0106045].
%%CITATION = QUANT-PH 0106045;%%





%\cite{Santangelo:2000fj}
\bibitem{Santangelo:2000fj}
E.~M.~Santangelo,
%``Evaluation of Casimir energies through spectral functions,''
Theor.\ Math.\ Phys.\  {\bf 131} (2002) 527 [Teor.\ Mat.\ Fiz.\
{\bf 131} (2002) 98] [arXiv:hep-th/0104025].
%%CITATION = HEP-TH 0104025;%%


%\cite{Milton:1999ge}
\bibitem{Milton:1999ge}
K.~A.~Milton,
%``The Casimir effect: Physical manifestations of zero-point energy,''
[arXiv:hep-th/9901011].
%%CITATION = HEP-TH 9901011;%%


\bibitem{O'Brien:pn}
K.~H.~O'Brien and C.~I.~Tan,
%``Modular Invariance Of Thermopartition Function And Global Phase Structure Of Heterotic String,''
Phys.\ Rev.\ D {\bf 36} (1987) 1184.
%%CITATION = PHRVA,D36,1184;%%


%\cite{McClain:1986id}
\bibitem{McClain:1986id}
B.~McClain and B.~D.~Roth,
%``Modular Invariance For Interacting Bosonic Strings At Finite Temperature,''
Commun.\ Math.\ Phys.\  {\bf 111} (1987) 539.
%%CITATION = CMPHA,111,539;%%






\bibitem{Ambjorn:1981xw}
J.~Ambjorn and S.~Wolfram,
%``Properties Of The Vacuum. 1. Mechanical And Thermodynamic,''
Annals Phys.\  {\bf 147} (1983) 1.
%%CITATION = APNYA,147,1;%%





\end{thebibliography}
\end{document}